\documentclass[a4paper,12pt]{article}
\usepackage{jheppub}	% jheppub includes hyperref, color, natbib, amsmath, amssymb, epsfig, graphicx
\usepackage[utf8]{inputenc}
\usepackage[english]{babel}
\usepackage{makeidx}
\usepackage{amsfonts}
\usepackage{enumerate}
\usepackage{mathrsfs}
\usepackage{tensor}
\usepackage[autostyle]{csquotes}
\usepackage{subfig}
\usepackage{simpler-wick}

\newdimen\tableauside\tableauside=1.0ex
\newdimen\tableaurule\tableaurule=0.4pt
\newdimen\tableaustep
\def\phantomhrule#1{\hbox{\vbox to0pt{\hrule height\tableaurule
width#1\vss}}}
\def\phantomvrule#1{\vbox{\hbox to0pt{\vrule width\tableaurule
height#1\hss}}}
\def\sqr{\vbox{%
  \phantomhrule\tableaustep
\hbox{\phantomvrule\tableaustep\kern\tableaustep\phantomvrule\tableaustep}%
  \hbox{\vbox{\phantomhrule\tableauside}\kern-\tableaurule}}}
\def\squares#1{\hbox{\count0=#1\noindent\loop\sqr
  \advance\count0 by-1 \ifnum\count0>0\repeat}}
\def\tableau#1{\vcenter{\offinterlineskip
  \tableaustep=\tableauside\advance\tableaustep by-\tableaurule
  \kern\normallineskip\hbox
    {\kern\normallineskip\vbox
      {\gettableau#1 0 }%
     \kern\normallineskip\kern\tableaurule}%
  \kern\normallineskip\kern\tableaurule}}
\def\gettableau#1 {\ifnum#1=0\let\next=\null\else
  \squares{#1}\let\next=\gettableau\fi\next}

\def\a{\alpha}		\def\b{\beta}				
						
				\def\l{\lambda}		
										
		\def\t{\tau}

						\def\L{\Lambda}

\def\be{\begin{equation}}
\def\ee{\end{equation}}
\def\bea{\begin{eqnarray}}
\def\eea{\end{eqnarray}}

\title{ On Horizon Molecules and Entropy in Causal Sets}

\author{D. Dou}

\affiliation{Dept of Physics, College of Exact Sciences, Hamma Lakhdar University, El Oued, Algeria.}
\affiliation{Lab of Linear Operators and Partial Diff Equations, Theory and Application, Hamma Lakhdar Univeristy.}

\emailAdd{dou-djamel@univ-eloued.dz}

\abstract{ We review the different proposals and attempts  to identify the ``horizon molecules"  that would give a kinematical estimation for the black hole entropy  in causal set theory. The proposals are presented  according to their chronological appearance in scientific   literature. The review is neither very technical nor  merely descriptive; it is aimed to provide the reader with a lucid introduction to the necessary concepts and mathematical background,  and  give him or her a  broad  view on the subject,  by focusing on the main technical and conceptual issues that summarize the progress made in the last two decades. 
 }

\keywords{Causal Sets, Quantum Gravity, Black Holes, Entropy, Horizon Molecules, Statistical Geometry.}

 \clearpage

\makeatletter
\gdef\@fpheader{}
\makeatother

\begin{document}
\maketitle

%\tableofcontents

\section{Introduction}

 Although the energy scale at which quantum effects on spacetime are expected to show up is well beyond the range of any foreseeable laboratory-based experiments, the theoretical consequences of quantum mechanics and general relativity  have been major reasons for studying quantum gravity and searching   for  a more fundamental structure of spacetime. Most importantly, the discovery of the close relationship between certain laws of black hole physics and the ordinary laws of  thermodynamics, on one hand, and the discovery of  the quantum induced radiation by black hole (BH),  on the other hand, appear to be two major pieces of
 a puzzle that fit together so perfectly that there can be little doubt that this ``fit”
 is of deep significance \cite{Wald:1999vt,  Sorkin:1997ja, Carlip:2014pma, Sorkin:2005qx, Bekenstein:1994bc}.
 
 Today, well into its fifth decade of the development, this merger remains intellectually stimulating and puzzling at once.

 One of the most puzzling aspects is the fact  black hole possesses  an entropy  equal to one quarter of its horizon area expressed in units of Planck area. And  in spite of five decades of intensive research,  debates and genuine    advances in different directions, especially within the context of string theory and 2+1 gravity \cite{Strominger:1996sh, Horowitz:1996qd, Maldacena:1997re, Carlip:2014pma, Carlip:1994gy}, see also \cite{Ashtekar:1997yu} for loop quantum gravity results, it is fair to say that the physical origin of this entropy and all questions accompanying the thermodynamic of BH are still lacking satisfactory answers, and the debate is far from being settled . In particular, it remains uncertain what ``degrees of freedom”  or microstates the entropy refers to, or  what unavailable information it quantifies.
 Moreover, it can be said that a well accepted criterion to select one approach out of the different  approaches to quantum gravity or a fundamental theory of nature is its success in solving black hole thermodynamics puzzles in  a satisfactory and general manner, in particular revealing  the statistical mechanics behind BH entropy . 
 
 It also is  generally believed that all the puzzles of the BH are not independent and will be solved once we really solve one of them. For this  and other reasons,   providing a controllable calculation of  BH entropy has been a prime target of all theories and proposals to quantum gravity.

 Indeed, in the current climate the
 role being played by BH thermodynamics in this connection looks more
 and more analogous to the role played historically by the thermodynamics
 of a box of gas  and black body radiation in revealing the underlying atomicity and quantum nature of
 everyday matter and radiation. This  analogy can be brought out more clearly by
 recalling some facts about thermodynamics in the presence of event horizons.

 A well accepted definition   of entropy  is as a measure of missing or ``unavailable” information about a physical system, and from this point of view, one would have
 to expect some amount of entropy to accompany an event horizon, since it is
 by definition an information hider par excellence, and therefore the BH entropy could be understood as a
 response of having an event horizon which hides information about a region of space
 time,  and here  the notion of entanglement entropy comes into play. This originates from the well known observation that an observer outside the horizon has
 no access to the degrees of freedom behind the horizon. For
 this reason the outside observer would describe the world
 with a reduced density matrix obtained by tracing out the
 inaccessible degrees of freedom behind the horizon. If the
 exterior modes and the external modes are correlated ``entangled’’ the resulting density operator is thermal even if the
 global state of the system is pure \cite{PhysRevLett.56.1885, Dowker:1994fi}.

 Now, what modes or missing information the BH entropy refers to generally remains  a mystery. Nevertheless, in the presence of a horizon,  in principle one should  associate to each quantum field  an ``entanglement
 entropy” that necessarily results from tracing out the interior modes of the field, given that these modes are necessarily correlated with the exterior ones. 
 In the continuum, this entanglement entropy
 turns out to be infinite, at least when calculated for a free field on a fixed
 background spacetime. However, if one imposes a short distance cutoff on
 the field degrees of freedom, one obtains instead a finite entropy; and if the
 cutoff is chosen around the Planck length then this entropy has the same
 order of magnitude as that of the horizon \cite{Bombelli:1986rw, Srednicki:1993im}. Based on this appealing result,
 there have been many speculations attributing the black hole entropy to the
 sum of all the entanglement entropies  of the fields in nature \cite{Bekenstein:1994bc}.  Whether or not the entanglement of
 quantum fields furnishes all of the entropy or part of it,
 contributions of this type must be present, and any consistent theory must provide for them in its thermodynamic
 accounting.
 
 It is not, of course,  the aim of this introduction to give an account of  the developments in different directions that have surrounded the entanglement entropy in connection with black holes, and reader is referred  for instance to \cite{Nishioka:2018khk}   and references therein. However; there is a growing consensus that entanglement entropy,  and in general quantum entanglement and holography,  will play a central role in revealing a  finer structure of spacetime and possibly leading to a radical revision of our perception of the universe.

At present, and  without having at hand a viable and more fundamental theory of spacetime, it is hard to expect a resolution of the problem of the divergence of entanglement entropy, which  is  very likely deeply linked   to  other issues of BH thermodynamics. Nevertheless, the finiteness of the BH entropy on one hand,  the behavior of  the entanglement  entropy in the continuum picture, on the other hand, seem to point directly towards an underlying discrete structure of spacetime.  The situation actually appears to be similar to that of an ordinary box of gas,
 where we know that, fundamentally, the finiteness of the entropy rests on
 the finiteness of the number of molecules, and to lesser extent on the discreteness of their quantum states. Indeed, at temperatures high enough to
 avoid quantum degeneracy, the entropy is, up to a logarithmic factor, merely
 the number of molecules composing the gas. The similarity with the BH becomes evident when we remember that the picture of the horizon as
 composed of discrete constituents gives a good account of the entropy if we
 suppose that each such constituent occupies roughly one unit of Planck area
 and carries roughly one bit of entropy 
 \cite{Sorkin:1997ja}.
 
  A proper statistical derivation along
 these lines would require a knowledge of the dynamics of these constituents,
 of course. However, in analogy with the gas, one may still anticipate that the
 horizon entropy can be estimated by counting suitable discrete structures,
 analogs of the gas molecules, without referring directly to their dynamics.
 Clearly, this type of estimation can succeed only if well defined discrete
 entities can be identified which are available to be counted. Within a continuum theory, it is hard to think of such entities. However, in causal set theory \cite{PhysRevLett.59.521}, the elements of the causal set
 serve as ``spacetime atoms”, and one can ask whether these elements, or some
 related structures, are suited to play the role of ``horizon molecules”.  
 
 The idea of considering a certain causal set structure  as a potential candidate for the horizon molecules   was first taken up in $1999$ using causal links. This  proposal  was  partially successful and gave promising results in 2 -dimensions. It was  subsequently followed by other proposals to refine it or look for more suitable definitions for the horizon molecules that would work in higher spacetime dimensions  .

 In this review, we  go through  the different horizon molecules proposals that emerged in the last two decades or so  within the causal set approach to quantum gravity.  The different proposals will be presented according to their chronological appearance in literature.   We therefore shall first focus  on the causal links proposal that appeared in \cite{Dou:1999fw, Dou:2003af}, which historically was the first proposal and so far seems to be the simplest one, and in spite of the fact that it  has turned out to be unsuccessful beyond 2-dimensions, this proposal remains pedagogically useful and conceptually stimulating    . As a consequence of the failure of the links proposal in higher dimensions,  other horizon molecules proposals were put forward  in subsequent and recent years aiming to succeed where the first proposal failed  \cite{Sarah,Barton:2019okw,Machet:2020uml}. These subsequent and recent  proposals   will then be reviewed, their main results will be reported and discussed. 
 
 This review is not intended to be a full comprehensive survey on this subject, however, we hope  that the material presented herein will offer the  beginner researcher in the subject,  or the interested theoretical physicist in general, an accessible introduction to the subject,  enough background, tools and concepts  that  enable him or her to   understand the above-mentioned efforts and developments  to identify the horizon molecules in causal set theory, and direct the reader to the still open issues.

 %The chapter is organized as follows. In the first section we give the necessary mathematical background about causal sets that is needed for carrying out the calculations within the different proposals.   The second s….
 
 %certain kind of “causal link” as one such structure
 %and we will show that the black hole entropy can be equated to the number
 %of such links crossing the horizon H in proximity to the hypersurface  for
 %which the entropy is sought. Moreover, almost all of these links will turn out
 %to be localized very near to H. In consequence, conditions deep inside the
 %black hole will become irrelevant to the counting, as indeed they must do if
 %any interpretation of the entropy in terms of “horizon degrees of freedom” is
 %to succeed 

\section{Background and Terminology}
In this section we give  the essential mathematical definitions and terminology  related to the causal set picture of spacetime.  We shall  limit ourselves to the necessary background  relevant to  this review. For more comprehensive and extensive introduction to causal set hypothesis we refer the reader to \cite{Meyer, Luca},  for a recent and broad review  with a fuller set of references  see \cite{Surya:2019ndm}. 

\hfill \break

  \textbf{Definition 1} A causal set (or a causet for short)  $\mathcal{C}$ is a set endowed with an order relation $\prec$
 satisfying the following axioms:
 \begin{enumerate}
 	\item 
 	Acyclic (antisymmetric): $\forall p, q \in \mathcal{C} , p \prec q ~~\text{and}~~ q \prec p  \Rightarrow p = q$ \ ,
 	\item
\text{Transitive}:$ \forall p, q, r \in \mathcal{C} , p \prec q \prec r \Rightarrow p \prec r$ \ ,

\item  Reflexive  : $\forall p \in \mathcal{C}, p \prec p$ \ ,
\item Locally finite: $\forall p, q \in \mathcal{C}, |I[p,q]|< \infty $, \text{where} $I[p,q]= \text{Fut}(p)\cap \text{Past}(q)$, $|.|$ stands for the cardinality of the set,   Fut and Past denote the future and the past of a given point,
$$
\text{Fut}(p)=\{ q\in \mathcal{C}| p\prec q, q\neq p \}
$$

$$
\text{Past}(p)=\{ q\in \mathcal{C}| q\prec p, q\neq p \} \ .
$$
 \end{enumerate}
 Notice here that the reflexivity axiom is a matter of convention and we could instead have
 used  the irreflexive convention.
 
 Fut$(p)$ and Past$(p)$ are to be compared with the notion of chronological future  and  past,  $I^+(p)$ and  $I^-(p)$, in continuum Lorentzian geometry. $ I[p,q]$ is referred to as the causal or order interval, the analogue of Alexandrov interval in the continuum.
 
 The discreteness of the causal set is encoded in the local finiteness  axiom. 
 
The acyclicity axiom ensures that causets do not have closed causal loops.
%\vspace{4}
 \hfill \break
 An important concept for the description of causets and that we shall frequently need is the \emph{Link}.
 
 \textbf{Definition 2}: Let $p$ and $q$ $\in \mathcal{C}$, $p\prec q$, $q\neq p$ . If $|I[p,q]|=0$, we say there is link between $p$ and $q$ and  write $p\prec\!\!\cdot \,  q$.
 
 The knowledge of all links is equivalent to knowledge of all relations among elements:
  $p \prec q $ iff there are elements $ q_1, q_2, , ......q_n$ such that $p \prec\!\! \cdot \,  q_1 \prec\!\! \cdot \, q_2 \prec\!\! \cdot ,  .....$ $\prec \!\!\cdot \, q_n \prec \!\!\cdot  \, q $. 
 Therefore  links are irreducible relations and in some sense are the building blocks of the  causet.
 %\hfill \break

 \textbf{Definition 3}: Let $\mathcal{C}' \subset \mathcal{C}$, $p \in \mathcal{C}'$ is said to be \emph{maximal} (resp.\emph{minimal}) in $\mathcal{C}'$ iff it is in the past (resp. future) of no other element in $\mathcal{C}'$.
 %\hfill \break
 
 An extended notion of maximality and minimality condition that will later be needed is the notion of maximal and minimal-but-$n$.

 \textbf{Definition 4}: Let $\mathcal{C}' \subset \mathcal{C}$, $p \in \mathcal{C}'$ is said to be \emph{maximal-but-$n$} (resp.\emph{minimal-but-$n$}) in $\mathcal{C}'$ iff it is in the past 
 (resp. future) of exactly  $n$  elements in $\mathcal{C}'$.

% \vspace{5}
\hfill \break
 The basic hypothesis of the causal set approach to quantum gravity is  that \emph{``spacetime, ultimately, is discrete and 
 its underlying structure is that of a locally finite, partial ordered set which continues
 to make sense even when the standard geometrical picture ceases to do so"}. The macroscopic spacetime continuum we experience must be recovered as an approximation to the causet. The causal set proposal can roughly be summarized in the following two points
 \begin{enumerate}
 	\item  Quantum Gravity is a quantum theory of causal sets.
 	\item A continuum spacetime $(\mathcal{M}, g)$ is an approximation of an underlying causal
 	set $C \sim (\mathcal{M}, g)$, where
 	
 	(a) Order $\sim$ Causal Order
 	
 	(b) Number $\sim$ Spacetime Volume
 \end{enumerate}
  
   Point or step (2) is not to be viewed as independent of step (1). Actually the quantum theory of causal set should dictate how the continuum picture  emerge as an approximation, and this could ultimately   involve a more sophisticated notion of approximation.  For instance, in view of the fact that not all causets admit a realization as spacetimes with a given dimension while respecting conditions (2a)  and (2b), the process by which   the continuum 4-d spacetime picture,  or that of  higher dimensional spacetimes  with compactified extra-dimensions, is reached may involve some sort of coarse-graining in which the manifold picture  would be  a scale
  dependent  approximation of the causal set. However, in the absence of a quantum dynamics of causet,  a systematic way of defining a coarse-graining that would fit automatically our expectations is yet to be discovered. Nevertheless, we may   use point (2) as a  stepping stone (given) to investigate possible kinematical consequences of the causet approach. In short and without expanding too much around this point, the intuitive idea at work here is that of a \emph{faithful embedding} which we define below.
  %\vspace{4} 
 
 \hfill \break
  \textbf{Definition 5}
  If $(\mathcal{M},g)$ is a $d$-dimensional Lorentzian manifold and $\mathcal{C}$ a
  causet, then a faithful embedding of $\mathcal{C}$ into $\mathcal{M}$ is an 
  injection map
  $f:\mathcal{C} \hookrightarrow\mathcal{M}$ of the causet into the manifold that satisfies the
  following requirements:

  \begin{enumerate}
  	\item 
  
 The causal relations induced by the embedding agree with those of
  $\mathcal{C}$ itself,

 i.e.   $x\prec y \Leftrightarrow f(x)\in J^{-}(f(y))$
  
   where
  $J^{-}(p)$ stands for the causal past of $p$ in $\mathcal{M}$;
  \item
   The embedded points are distributed uniformly at density $\varrho_c=l_c^{-d}$ with respect to the spacetime volume measure of $(\mathcal{M},g)$.
    
\item
   The characteristic length over which the geometry
  varies appreciably is everywhere much greater than the mean spacing
  between the embedded points. 
  \end{enumerate} 

 $l_c$ is referred to as the discreteness scale.

  When these conditions are satisfied,
  the spacetime $(\mathcal{M},g)$ is said to be a
  continuum approximation to $\mathcal{C}$ and we write $\mathcal{C} \sim (\mathcal{M},g). $
  
  To ensure covariance the above embedding is realized by  randomly sprinkling in points until the required density is reached. Therefore from the point of view of $\mathcal{M}$ the causet resembles a ``random
  lattice'', e.g  ``a regular" lattice cannot do the job since
 it is not uniform in all frames or coordinate systems.

  % ِA  natural choice for the map $f$ to achieve this
   A natural choice for obtaining or  creating a faithfully embedded causet is via a 
   Poisson point process; under which the 
   probability to find $n$ elements in a spacetime region of volume $V$ is given by
  
 \begin{equation}
 	(\varrho_c V)^{n}\frac{e^{-\varrho_c V}}{n!} \ .
 \end{equation}
  This  makes $f(\mathcal{C})$ a random causet and thereby any function $F:\mathcal{C} \rightarrow \mathbb{R}$ is a random variable. 
  
  For more detailed discussion of the issue of faithful embedding and the probabilistic nature of the process we refer the reader to \cite{Surya:2019ndm} and references therein.

  \section{Horizons molecules as causal links }
  As discussed in the introduction, the expectation is that the BH entropy can be understood  as
  entanglement in a sufficiently generalized sense, and we may hope to
  estimate its leading behavior by counting suitable
  discrete structures that measure the potential entanglement in some
  way between in-outside discrete structures. Moreover, and owing to the fact that the entropy  essentially measures  the horizon area in Planck units,  the problem is reduced to  coming up with this measure in the causal set picture. 
  
  It  is worthy of note here that it seems far from obvious that such structures must exist.  If
  they do, then they provide a relatively simple order theoretic measure
  of the area of a cross section of a null surface, and, unlike what
  one's Euclidean intuition might suggest, it is known that such measures
  are not easy to come by.  For example, no one knows such a measure of
  spacelike distance between two sprinkled points that works in general, though some progress has been made in  such
   Minkowski
  spacetime \cite{Rideout_2009}. 
  
  It follows from the above discussion that  a natural and the simplest candidate for the structure we seek is a \emph{link} crossing the horizon.  Indeed, we may think heuristically of
  ``information flowing along links'' and producing entanglement when it
  flows across the horizon during the course of the causet's growth (or
  ``time development'').  Since links are irreducible causal relations (in
  some sense the building blocks of the causet), it seems natural that by
  counting links between elements that lie outside the horizon and
  elements that lie inside, one would measure the degree of entanglement
  between the two regions.  Equally, it seems natural that the number of
  such causal links if supplemented with extra conditions   might turn out to be proportional to the horizon area and play the role of the Horizon molecules.
  
  %Historically the first proposal to use causal links as horizon molecules was put forward in cite. This proposal seems to work in 2-d, however it fails in higher dimension due to IR divergences, and the way it failed led recently  to the construction of different structures .
  
  In what follows we discuss with some detail the  links proposal for horizon molecules and its applications in different 1+1 geometrical setups. 
  
 \subsection{The general Setup } 
 
 Let us consider a causet $\mathcal{C}$ obtained via  Poisson random sprinkling in a black hole
 background  $\mathcal{M} $ with density $\varrho_c$, so this causal set is faithfully embeddable in this geometrical background by definition. Let $\mathcal{H}$ be a BH horizon and let $\Sigma$  be an achronal hypersurface intersecting the horizon, Figure 1. 
 
 \begin{figure}
 	\begin{center}
 		\includegraphics[height=4in,width=7in,angle=0]{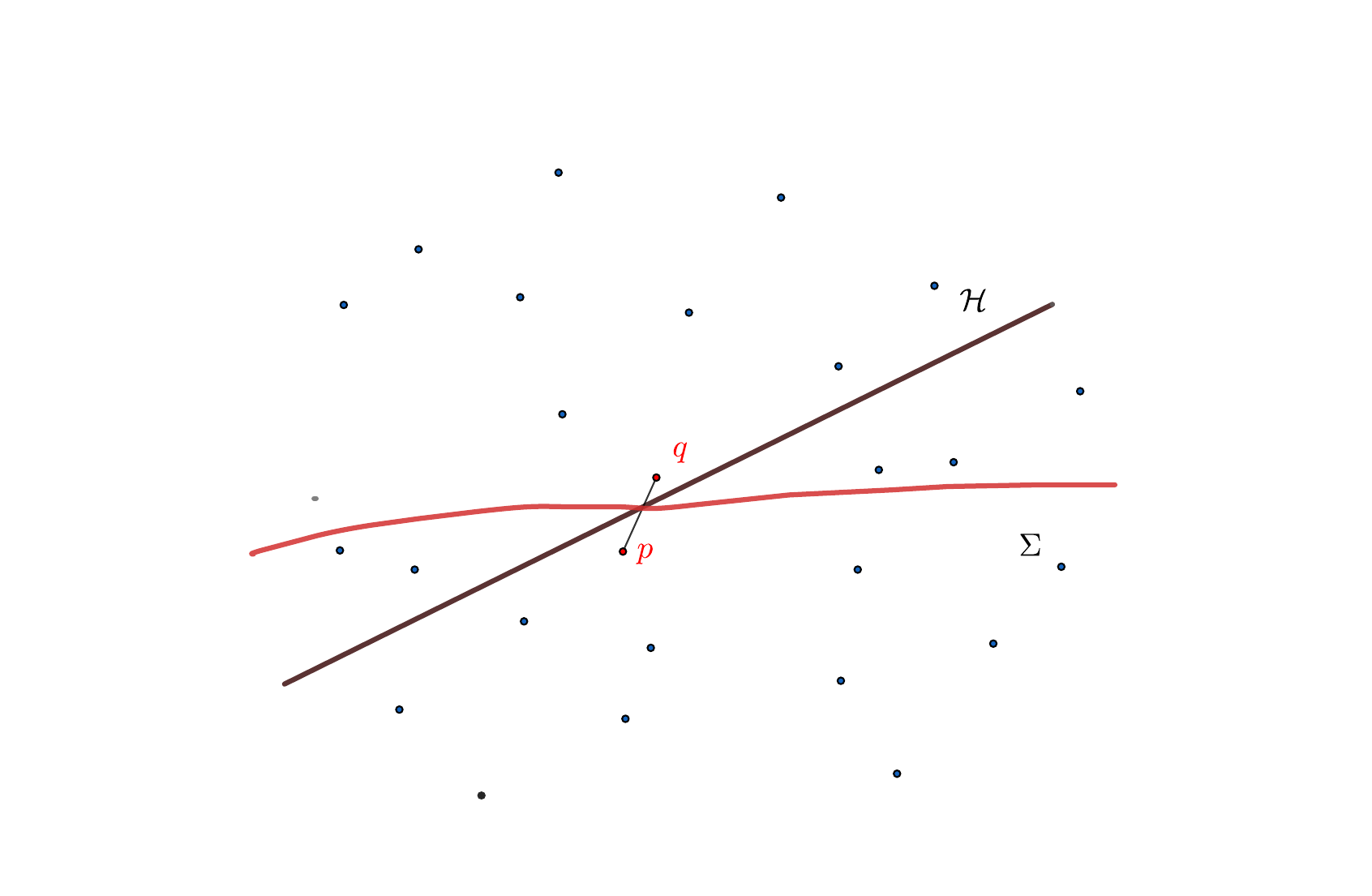}
 		\caption{A typical geometrical setting showing a typical causal link crossing the horizon.}
 	\end{center}
 \end{figure}
 
 The goal is to come up with a measure of the area  of the resulting cross section between $\mathcal{H}$ and $\Sigma$, which in turn would measure the horizon entropy and define \emph{Horizon molecules}.
 
 A natural and intuitive candidate for such molecules is to take them  made of pairs of points $(p,q)$,  with $p$ lying  outside the black hole and to the past of $\Sigma$, while $q$ is inside the black hole and to the future of  $\Sigma$, and $p\prec\!\!\cdot \ q$, i.e. $p\prec  q$ is a link   .  
 
 If no further conditions are imposed on $p$ and $q$, the expected number of such  links can easily be shown to diverge. 
 
To see what conditions must be imposed on the pairs $ (p,q)$, let us  remember that, intuitively, what we are trying to estimate 	is not the total sum of all ``lost information" but only that corresponding ``to a given time", meaning in the vicinity of the given hypersurface $\Sigma$. Hence, to  associate the same causal link with more than one hypersurface would be to `` overcount" it in forming our estimate, and it is this   overcounting that seems to be the source of the above mentioned divergence.  Therefore further conditions are needed to be imposed to give a definition of the horizon molecules  which is truly proper to $\Sigma$ rather to some  earlier or later hypersurface. Several possibilities suggest themselves for this purpose, but none seems to be clearly best, as the end result (the leading order) will be shown to be insensitive to which choice one makes. Below we pick up a specific choice or definition of horizon molecules, which will be referred to as the ``\emph{causal links proposal}",  and the general issue will be discussed further in subsection  \ref{maxmin} . Actually working out explicitly with this particular choice, seeing its success in 1+1 and failure in higher dimensions, due to IR divergence,  will be instructive for the reader to conceive the motivations behind  the   re-definitions of the horizon molecules   that subsequently departed from the original links proposal.

 \hfill \break

\textbf{The causal links proposal (Dou-Sorkin 1999)}: A horizon molecule with respect to  a given hypersurface $\Sigma$ is  a pair $(p,q)$ satisfying the following conditions

\begin{enumerate}
	\item $p \in I^-(\Sigma ) \cap I^-(\mathcal{H})$
	\item $q \in I^+(\Sigma ) \cap I^+(\mathcal{H})$
	\item $|I[p,q]|=0$, i.e $p\prec q$ is a link.
	\item $p$ is maximal in $I^-(\Sigma ) \cap I^-(\mathcal{H})$ and $q$ is minimal in $I^+(\mathcal{H})$ .
\end{enumerate}
	
	The $4^{th}$ condition may seem asymmetric, as one would have expected   symmetric Max and Min conditions between $p$ and $q$ to be  more natural, however, the reason that 
  we do not impose a similar condition on $q$ is because this  would give zero
  for a null hypersurface case, but the result should agree for null or spacelike if both intersect the
  horizon in the same time, moreover for stationary black the results should agree in
  all cases.

Before we move on, we draw the reader's  attention that throughout  this section and the next one $p$ will stand for  points in $I^-(\Sigma ) \cap I^-(\mathcal{H})$ and $q$  for the ones  in  $I^+(\Sigma ) \cap I^+(\mathcal{H})$.

Let us now see how to count the expected  number of these horizon molecules  by reducing it to the calculation of an integral over the manifold    $\mathcal{M} $.

%As  we mentioned earlier any function $F:\mathcal{C} \rightarrow \mathbb{R}$ is a random variable. 

Remember  that the probability of finding or sprinkling $n$ points in some region of spacetime, $\mathcal{R}$, is given by the Poisson distribution 

$$
P(n,\mathcal{R} )=\frac{(\varrho_c\text{vol}(\mathcal{R}))^n}{n!} e^{-\varrho_c\text{vol}(\mathcal{R})} \ ,
$$

where $\text{vol}(\mathcal{R})$ is the spacetime volume of $\mathcal{R}$.
	
Consider first an infinitesimal region $ \Delta \mathcal{R}$, the probability of sprinkling a single point in it is  follows from 

\begin{equation}
P(1,\Delta\mathcal{R})\approx \varrho_c \text{vol}\Delta\mathcal{R}\equiv  \varrho_c \Delta V \ . 
\end{equation} 

Consider now two infinitesimal regions $  \Delta \mathcal{R}_p \in I^+(\Sigma ) \cap I^+(\mathcal{H}) $ and $  \Delta \mathcal{R}_q\in  I^-(\Sigma ) \cap I^-(\mathcal{H})$. The probablity of having  a pair of points $(p,q)$ with $p \in I^+(\Sigma ) \cap I^+(\mathcal{H})$ and $q \in  I^-(\Sigma ) \cap I^-(\mathcal{H})$ sprinkled in $  \Delta \mathcal{R}_p$ and $  \Delta \mathcal{R}_q$ resp. is given by

\begin{equation} 
P(p,q | \Delta \mathcal{R}_p, \Delta \mathcal{R}_q)=\varrho_c\Delta V_p \varrho_c\Delta V_q \ .
\end{equation}
If we further require the relation between $p$ and $q$ to be a link then the Alexandrov interval $A(p,q)$ between $p$ and $q$   must contain no point and therefore the  probability becomes

\begin{equation} 
	P(p\prec \!\!\cdot  \, q | \Delta \mathcal{R}_p, \Delta \mathcal{R}_q)=  P(0,\text{vol}(A(p,q))\varrho_c^2\Delta V_p \Delta V_q= \varrho_c^2e^{-\varrho_c\text{vol}(A(p,q))}\Delta V_p\Delta V_q \ .
\end{equation}

 In addition to the link condition Max and Min conditions must be imposed on $p$ and $q$. The Max and Min conditions are just statements about an extra region in $\mathcal{M}$ being empty, with no sprinkled points. If we denote by $\mathcal{R }(p,q)$ the region  resulting from the union of $A(p,q)$, $I^+(p)\cap I^-(\Sigma)\cap I^-(\mathcal{H})$ and $I^-(q)\cap I^-(\mathcal{H})$  the probability for the above link to become  a horizon molecule  reduces to

 \begin{equation}\label{HM1} 
 	P(\mathbf{H}(p,q) ;\Delta \mathcal{R}_p, \Delta \mathcal{R}_q) = \varrho_c^2e^{-\varrho_cV(p,q)}\Delta V_p \Delta V_q \ ,
 \end{equation}

where $V(p,q) = \text{vol}(\mathcal{R }(p,q))$.

To count the expected number of horizon molecules we remember that the existence of  horizon molecule is a random variable generated by  a function   whose value is $1$ if the horizon molecule conditions are fulfilled  and $0$ otherwise. With this in mind, it follows that expected number of horizon molecules is obtained by summing in (\ref{HM1}) over all $p \in I^-(\Sigma ) \cap I^-(\mathcal{H})$ and $q \in I^+(\Sigma ) \cap I^+(\mathcal{H})$ in the limit $\Delta V_p$ and $ \Delta V_q$ go to zero.  In this limit the sums are replaced by integrals over the domain of $p$ and $q$ to obtain the following final expression for the expected number of horizon molecules

\begin{equation}\label{HMF}
	<\mathbf{H}_{link}>= \varrho_c^2\int_{I^-(\Sigma ) \cap I^-(\mathcal{H})} dV_p\int_{I^+(\Sigma ) \cap I^+(\mathcal{H})}    dV_q ~e^{-\varrho_cV(p,q)} \ .
\end{equation}
 
For a more systematic derivation of the above integral formula see \cite{Sarah} .

 For  horizon molecules as such to be successful, one has to show that in the limit of large density, or $l_c$ is much smaller than the geometrical length scales of the setting, 	$<\mathbf{H}_{link}> $ has the asymptotic form
 
 \be
\varrho_c^{\frac{d-2}{d}} <\mathbf{H}_{link}> = a^{(d)}\int_{\mathcal{J}} dV_\mathcal{J}+\cdots \ ,
 \ee 
  where the dots refer to terms vanishing   in the continuum limit. $\mathcal{J}:= \Sigma \cap \mathcal{H}$ and $dV_\mathcal{J}$ is the surface measure on $\mathcal{J}$. $a^{(d)}$ is constant that depends on the dimension of the spacetime but, in principle, not on the nature of $\Sigma$, null or spacelike. In two dimensions the leading term in $<\mathbf{H}_{link}>$ should  be just a constant.

\subsection{Horizon molecules and the area law in 2-dimensions}\label{link}

  Ideally one would have used (\ref{HMF}) to the evaluate the expected number of horizon molecules, $<\mathbf{H}_{links}>$, in a full four dimensional BH background, e.g Schwarschild BH, however, historically and for technical reasons (difficulties)
 a simplified version was  first worked out. This consisted in  considering  a `` dimensionally reduced" two dimensional  metric instead of the true four dimensional one.  The hope was twofold; it would first be a warm up exercise for a more realistic four dimensional BH; second the establishment of the area law in 2-d models would give strong evidence for the validity of this proposal in the full four dimensional case. Stated differently, the four-dimensional answer would differ from the two-dimensional one only by a fixed proportionality coefficient of order one, together with a factor of the horizon area. 
 
 Now, although the above defined horizon molecules  proposal did not work  beyond $2$-d, in contrast to what   had first been hoped, due to IR divergences, the establishment of the area law in 2-d using the above defined horizon molecules makes the calculation worth  discussing . Beside this obvious reason, it will be seen that in $1+1$ the resulting expected number of links seems to exhibit some interesting features: a sort of universality, giving exactly the same answer for two different geometrical backgrounds, in equilibrium and far from equilibrium, and remaining finite in the strict continuum limit, $\varrho_c \rightarrow  \infty$.   
 
In the sequel  two cases will explicitly be worked out, a 2-d reduced Schwarschild geometry and collapsing null shell. We  shall set $\varrho_c=1$ in all 2-d models discussed in this section; because the leading term is  a dimensionless constant and the subleading ones are easy to  express and control in these units.

\subsection{An equilibrium black hole: 2-d reduced model}\label{SBH}
Consider a dimensionally reduced  Schwarzschild spacetime obtained from the realistic
$4$-dimensional BH spacetime, outside a collapsing spherically symmetric star, by identifying each $2$-sphere $S^2$ to a point. The
resulting two dimensional spacetime has exactly the same causal structure as the
S-sector of the 4-dimensional one. The Penrose diagram for this spacetime is depicted in Figure 2 . For simplicity  the presence of the collapse has been ignored; this of course will not
change the argument, since the detail of the collapse should be irrelevant, or one can choose the hypersurface to intersect the horizon far from the collapse
and the result will not be affected by the presence of collapse.

The line element of the resulting spacetime is obtained by omitting the angular coordinates from the four dimensional line element, namely
\begin{equation}\label{le}
	d^2s =-\frac{4a^3}{r}e^{-r/a}du dv \ ,
\end{equation}

where $a=2M$ is the radius of the BH and $u$ and $v$ are the usual Kruskal-Szekeres coordinates, with $r$ defined implicitly by the equation 
\begin{equation}\label{le2}
	uv =(1-\frac{r}{a}) e^{r/a} \ .
\end{equation}
The  associated volume element is
\begin{equation}
	dV =\sqrt{-g} du dv=\frac{2a^3}{r}e^{-r/a}du dv \ .
\end{equation}

Our signs convention is such that $u\sim t-r$, $v\sim t+r$, and the horizon $\mathcal{H}$ coincides with $u=0$.
\begin{figure}
	\begin{center}
		\includegraphics[height=3.3in,width=5in,angle=10]{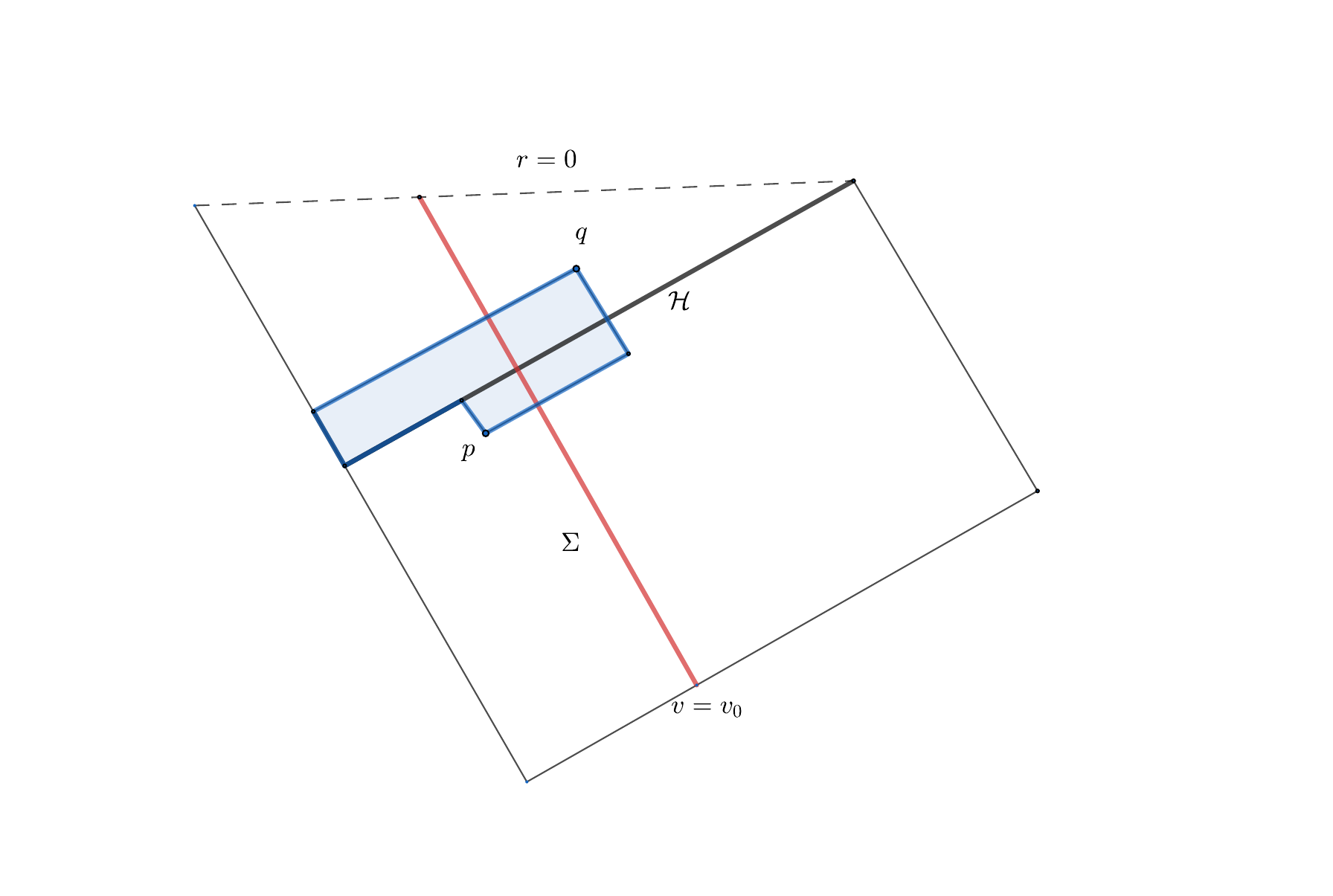}
		\caption{An equilibrium BH obtained from real the 4-d Shwarszchild BH by dimensional reduction, keeping only the radial section. The shaded region is required to be free from any sprinkled points and with  volume $V(p,q)$. }
	\end{center}
\label{Figure 2}
\end{figure}
Let now $\Sigma$ be an ingoing null hypersurface defined by the equation $v=v_0$ . The shaded region depicted in Figure 2 is the region $\mathcal{R }(p,q)$ with no sprinkled point,  its $ V(p,q)$ volume can  readily  be evaluated using (\ref{le2}) 
\begin{equation}\label{vol1}
	V=a^2+r_{pq}^2 -r_{pp}^2 -r_{qq}^2 \ ,
\end{equation}
where we have introduced the following notation

\begin{equation}
 u_i v_j = \left(1-\frac{r_{ij}}{a}\right) e^{r_{ij}/a} \ .
\end{equation}

Let us note that in two dimension and for a null $\Sigma$ the maximality condition on $p$ is actually redundant and insured by the link condition, but it would be needed with spaclike $\Sigma$.

  Using (\ref{HMF}) and (\ref{vol1}), the expected number of horzion molecules  is given by

  \begin{equation}
  	<\mathbf{H}_{link}>
   =(2a^3)^{2}
  	\int_{0}^{v_{0}} dv_{p}
  	\int_{-\infty}^{0} du_{p}
  	\int_{v_{0}}^{\infty} dv_{q}
  	\int_{0}^{1/v_{q}} du_{q}
  	\frac {e^{-r_{pp}/a - r_{qq}/a}} {r_{pp} r_{qq}} \; e^{-V} \ .
  \end{equation}

%ِِA change of integration variables 

A change of integration variables
from $(u_p,v_p,u_q,v_q)$ to $(r_{pp},  r(u_p,v_0)\equiv r_{p0},r_{pq},r_{qq})$,
followed by the notational substitutions
$x=r_{pq}$, $y=r_{p0}$, $z=r_{pp}$,  reduces $	<\mathbf{H}_{link}>$  to the form,
$$
<\mathbf{H}_{link}> = 4 \, I(a) \, J(a)  \ ,
$$
where

\begin{equation}
	I(a) 
	=
	\int_{a}^{\infty }dx \frac{x}{x-a} e^{-x^{2}} 
	\int_{a}^{x}dy \frac{y}{y-a} 
	\int_{a}^{y} e^{z^{2}}dz \ ,
\end{equation}

and
\be
J(a) = e^{-a^{2}} \int_{0}^{a} e^{r_{qq}^2} dr_{qq} \ . 	
\ee

It is worth noting here that the initial explicit dependence of $<\mathbf{H}_{link}>$ on $ v_0$ has disappeared, reflecting the stationarity of the black hole.

 Now, inasmuch as comparison with the Bekenstein-Hawking entropy is
meaningful only for macroscopic black holes, it is natural to assume
that $a\gg 1$, and under this condition, $I(a)$ can be shown  to have the following asymptotic behavior  \cite{Dou:1999fw}:
$$
I(a) = {\pi^2 \over 12} \; a + \mathcal{O}\left({1 \over a}\right)  \ .
$$
On the other hand it is not difficult to see that
$$
J(a) = \frac{1}{2a} + \mathcal{O}\left( {1 \over a^3} \right)  \ .
$$
%%       e^{-a^{2}} \int_{0}^{a} e^{r_{yy}^{2}} dr_{yy} 
Putting everything together, we end up with
\begin{equation}\label{Area2-1}
<\mathbf{H}_{link}> = \frac{\pi^2}{6} + \mathcal{O}\left( \frac{1}{a^2} \right) \ .  
\end{equation}

As the intersection of $\Sigma$ and $ \mathcal{H}$ in two dimension  is just a point, the area law, if finite, should   naturally turn out to be a pure number, therefore (\ref{Area2-1}), or the expected number of horizon molecules, is  proportional to the area  of the horizon in $1+1$.

Some remarks about the above derivation of the area law in 2-d using this horizon molecules proposal are in order.

The first remark concerns the locations of the pairs forming the  molecules  that give the dominant  contribution to $ <\mathbf{H}_{link}>$. It is easy to see that  the dominant contribution to the integral $J(a)$  plainly comes from $r_{qq}\approx{a}$, but since
$r_{qq}$ is the radial coordinate $r$ of sprinkled point $q$, and since
$r=a$ is the horizon, this implies that $q$ resides near the horizon.
Similarly,  an inspection of the integral $I(a)$  shows that
the dominant contribution to the integral $I(a)$  comes as well from
$z\approx y \approx {a}$, which, since $z=r_{pp}$ and $y=r_{p0}$, implies in turn that sprinkled
point $p$ resides near the horizon as well \cite{Dou:1999fw}.  Consequently this counting
can be said to be controlled by the near horizon geometry. 

It should be noted too that  from  the unboundedness of the region $I^+(\Sigma ) \cap I^+(\mathcal{H})$  and the finitness of $<\mathbf{H}_{link}>$, we can infer that points $q$ sitting arbitrarily close to the horizon but far from the $\Sigma \cap \mathcal{H}$ cannot continue to contribute indefinitely to $<\mathbf{H}_{link}>$. Moreover, the fact that $<\mathbf{H}_{link}>$ turns out to be just a pure number strongly suggests  that the pairs which give the dominant contribution are not only residing near the horizon but are   hovering near $\Sigma \cap \mathcal{H} $ too   .

It is interesting to look at this  result and its features from another point of view. If we inspect the integral $I(a)$ we note that what makes the near horizon molecules special is the vanishing of the denominators in $I(a)$ when the dummy integration 
 variables $x$ and $y$ tend to $a$.  To the extent that it is
this divergence which makes the horizon such a strong source for the
links, and here we may be reminded of the analogous fact that the strong redshift
in the vicinity of the horizon allows modes of arbitrarily high (local)
frequency to contribute to the entanglement entropy without influencing
the energy as seen from infinity.  Notice also that the clustering of
$p$ and $q$ near the horizon is not simply a consequence of the
maximality and minimality conditions we imposed on them.  For instance,
pairs $(p,q)$ sitting arbitrarily close to the hypersurface $\Sigma$,
with $q$ arbitrarily close to the horizon, still do not contribute to the
leading term in $I(a)$ if $q$ is far from the horizon, namely with
coordinate $\left|u_{p}\right|\gg 1$.
 
\subsection{A black hole far from equilibrium: 2-reduced collapsing null matter}

We now turn to another case which, though still spherically symmetric, is very far from equilibrium, namely that of a spherically collapsing null shell of matter with stress energy tensor
given by
$$
T_{vv}=\frac{M\delta(b-v)}{4\pi r^2} \ ,
$$

and the other components are identically zero. 

The collapsing shell forms a Schwarzschild BH. The Penrose diagram for the resulting spacetime (after dimensional reduction $S^2 \rightarrow$ point) is shown in Figure 3. Let the
shell sweep out the world sheet $v=b$ and let us choose for our
hypersurface $\Sigma$ a second ingoing null surface defined by $v=a$,
with $a<b$ so that $\Sigma$ lies wholly in the flat region.  Here $a$  
is of course generally different from $a$ defined  in the Schwarzschild case, $u$
and $v$ are null coordinates, chosen so that the horizon first forms at
$u=v=0$ and normalized for convenience such that the line element in the flat region is given by
$$
ds^2 = - 2 du dv + r^2 d\Omega^2 \ .
$$
Since our interest is again in macroscopic black holes, we will assume as
before that the horizon radius at $\Sigma \cap \mathcal{H} $  is large in units such that $\varrho_c=1$, which amounts to  $a{\gg}1$;
and to simplify matters further, we will also restrict ourselves to a
time well before the infalling matter arrives
(as judged in the center of mass frame).
One thus has the double inequality, $b{\gg}a{\gg}1$.  
Once again,  the calculation will be performed  for the two dimensional
radial section rather than the full four dimensional spacetime.

\begin{figure}
	\begin{center}
		\includegraphics[height=4.1in,width=7in,angle=0]{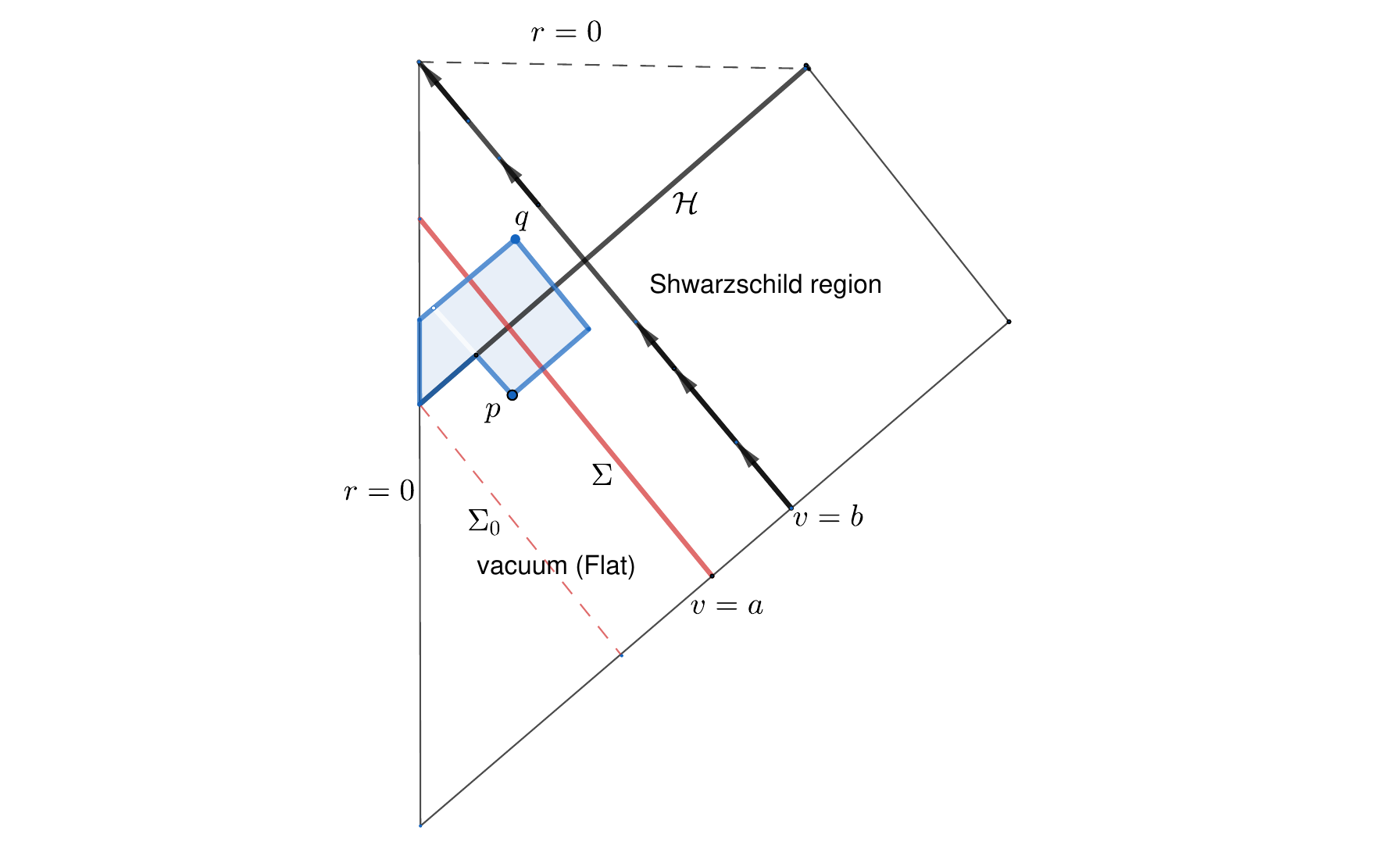}
		\caption{A non-stationary BH. The region to the past of the world line of the infalling matter is flat space with an expanding event horizon, whereas the one to its future is Shwarzschild region. We have depicted an extra null hypersurface $\Sigma_0$ for later reference.}
	\end{center}
\end{figure}

Since we are assuming that the infalling matter is far to the future
of the hypersurface $\Sigma$, points $q$ sprinkled into that region
should not contribute significantly when our minimality and link
conditions are taken into account.  For this reason, we shall, for
convenience, restrict the counting to pairs $(p,q)$ with $v_q<b$.

Using the definition of the horizon molecules we introduced above, one
obtains for the expected number of horizon molecules

\begin{equation}\label{integnull}
<\mathbf{H}_{links}>
=
\int_{a}^{b} dv_{q} 
\int_{0}^{v_{q}} du_{q} 
\int_{-\infty}^{0} du_{p}
\int_{0}^{a} dv_{p}  
e^{-V} \ ,
\end{equation}

where $V=u_{q}v_{q}-u_{p} (v_{q}-v_{p}) - u_{q}^{2}/2$, the volume of shaded region in Figure 3. 

Note here that the contribution of the points $p$ with $v_p<0$, i.e. to the past of $\Sigma_0$,  has been ignored; we will  return to its justification below.

The integration over $v_p$ and $u_p$ is easy to perform, followed by change of variables, $x=v_q, y=v_q-u_q$, we end up with
 \begin{equation}
 	<\mathbf{H}_{link}>
 	=
 	\int_{a}^{b}  \ln (\frac{x}{x-a}) e^{-x^2/2}dx
 	\int_{0}^{x} e^{y^2/2} dy \ .  
  \end{equation}
At this stage it is not difficult to show that the leading behavior of this integral for
 large $a$ is given by

 	\be\label{Hlinksnnull}
 	<\mathbf{H}_{link}>
 	=
 	{\pi^2 \over 6} - l  \left({a \over b}\right) + \mathcal{O}(1/a^2)  \ ,
 	\ee
 	where $l(x)\equiv \sum_{k=1}^\infty{x}^k/k^2$, a convergent series that
 	vanishes in the limit $x\to{}0$.  
 	
 	Originally the correction to the leading term in (\ref{Hlinksnnull}) were set to be of the order of $1/a$ in \cite{Dou:1999fw} and \cite{Dou:2003af}, but a careful repetition of the calculation due to Marr showed that the correction is of the order $1/a^2$ \cite{Sarah}. 
 	
 	Since we have assumed that $a{\ll}b$, we can write this more simply as
 \be\label{Area-2-null}
 <\mathbf{H}_{link}> = {\pi^2 \over 6} + \mathcal{O}(a/b) + \mathcal{O}(1/a^2) \ . 
 \ee

 	Notice that the presence of a negative contribution like $-l(a/b)$ was
 	to be expected, since we have omitted to count molecules that extend past
 	the shell into the Schwarzschild region.  For $\Sigma$ near to the
 	shell, one obviously should not neglect such links, and this counting is
 	incomplete. However, if the collapse is pushed far away from $\Sigma$, in particular to future infinity, we can safely  restrict the counting to the flat region  without worrying about the presence of the Schawrzschild region and therefore reducing the problem (even in higher dimension) to a counting in flat background geometry. 
 	
 	Now, what is  striking about the above result is is the occurrence of
 	the same numerical coefficient ${\pi^2}/{6}$ in both (\ref{Area-2-null}) and
 	(\ref{Area2-1}).  This agreement seems at first sight to furnish a nontrivial consistency check
 	of the suggestion that one can attribute the horizon entropy to the horizon molecules made of 
 	``causal links'' crossing it.

 As mentioned above, in writing  (\ref{Hlinksnnull}) we  implicitly ignored the contribution  of pairs $(p,q) $ with negative $v_p$. No justification for this was given in \cite{Dou:1999fw} nor in \cite{Dou:2003af}. However, this point was raised and briefly discussed by Marr in \cite{Sarah}.
 
   It is  easy to write an integral formula for this type of contribution, and maybe compute it, however, it is not difficult to argue that it should not be considered as part of the horizon molecules associated with $\Sigma\cap \mathcal{H}$. This kind of contribution counts the expected number of horizon molecules associated with a hypersurface $\Sigma_0$, Figure 3, which is not intersecting the horizon, or they occur before the horizon formation. Therefore they must be taken as sort of random statistical fluctuations  extraneous  and  not genuine horizon molecules associated to $\Sigma$. Actually, if we remember that the geometrical setting we are using is $2$-d reduced of a $4$-dimensional one, this extra contribution  would turn out to be just of order one in genuine $4$-dimensional counting, thus a negligible  fluctuation around the mean value.

 	\subsection{On the Min/Max conditions}\label{maxmin}
 	As we briefly discussed before picking up the particular choice for the``Max/Min" conditions we adopted in the definition  the causal links proposal,  this choice did not seem unique or particularly sacred and other variants were possible. Of course, one  must be careful not to
 	use something like ``$q$ minimal in $I^+(\Sigma)$'', which would drive
 	$<\mathbf{H}_{link}>$ to zero in the limit of null $\Sigma$, but this does not
 	rule out, for example, a condition like ``$p$ maximal in
 	$I^-(\Sigma)$''.
 	
 	Let us note that it turns out that there are at least two  variants of the ``Max/Min" condition that seem to be equivalent, as long as the leading term is concerned. These variants are

 	\textbf{Variant 1}:
 		$p$ is Max in $I^-(\Sigma )$ and $q$ is Min in $I^+(\mathcal{H})\cap I^+(\Sigma ) $ .

 	\textbf{Variant 2}:
 	$p$ is Max in $I^-(\Sigma )\cap I^-(\Sigma ) $ and $q$ is Min in $I^+(\mathcal{H})\cap I^+(\Sigma ) $ .
 
 	If we consider for instance the first variant, it is easy to show  that the resulting expected number of horizon molecules has the same asymptotic behavior as the one resulting from the original causal links  proposal\cite{Dou:1999fw}, namely
 		
 	$$
 <\mathbf{H}_{link}> = {\pi^2 \over 6}  + O(1/a^2) \ .	
 	$$
 	Thus,  for this variant at least, one obtains exactly the same numerical answer as the original proposal we started with. As for the second variant,  although it has  not been worked  out explicitly,  we do not expect the slight change in the  volume $V(p,q)$ should alter the leading term.
 	\begin{figure}
 		\begin{center}
 			\includegraphics[height=3in,width=5in,angle=0]{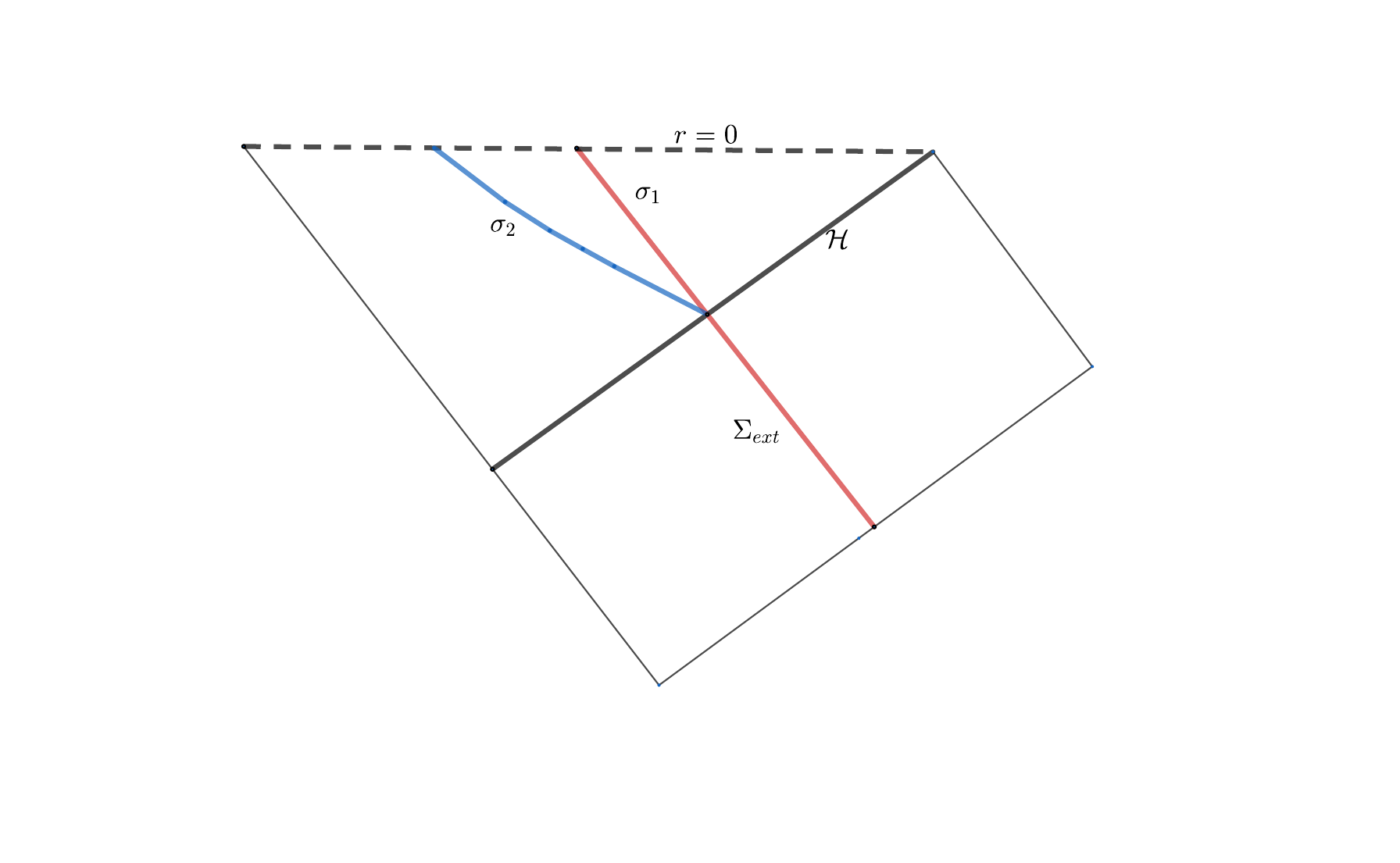}
 			\caption{Two continuations of a hypersurface to the interior region.}
 		\end{center}
 	\end{figure}

 	Another related feature the links counting must have if it is to yield the
 	horizon area is that, within reason, the expected number of horizon molecules should
 	depend only on the intersection $\mathcal{H}\cap\Sigma$, and not on how the
 	surface $\Sigma$ is prolonged outside or (especially) inside the
 	horizon $\mathcal{H}$.  For example one should get the same answer for both of
 	the continuations shown in Figure 4. The case where the difference is
 	confined to the interior black hole region is of particular significance
 	for the entanglement interpretation of horizon entropy, since such a
 	difference cannot, by definition, influence the effective density
 	operator for the external portion of $\Sigma$ (at least to the extent
 	that unitary quantum field theory is a good guide). For instance, we note that the volume  $V(p,q)$ needed to insure $p$ maximal in $I^-(\mathcal{H} )\cap I^-(\Sigma ) $ and  $q$  be minimal in $I^{+}(\mathcal{H})$ is the same for both $\Sigma_{ext}\cup \sigma_1$ and  $\Sigma_{ext}\cup \sigma_2$, therefore from this perspective   the  definition we have  so far adopted  seems to have  advantage over the other two variants,  at least in the case of null $\Sigma$.  However, in view of the fact that the leading order is controlled by contributions coming from links residing near the horizon, we expect the different variants to have the same leading behaviors no matter how $\Sigma $ is prolonged inside or outside the horizon.  Indeed, the issue of which Max/Min condition is favored cannot be settled nor properly discussed  unless we settle the central issue of how to define the horizon molecules in a way that works in higher dimensions and for both types of hypersurfaces, null and spacelike.

 	\subsection{The spacelike hypersurface in 2-d reduced }\label{spacelike}
 	
 	So far the counting of the horizon molecules has been restricted to null hypersurfaces in 2-d reduced black hole geometries. However, no proposal can be considered fully successful, even in two dimensions, unless it correctly reproduces the same result for both null and spacelike hypersurfaces. 
 	
In this section, we  look at this issue by discussing  the previous links counting when $\Sigma$ is  a spacelike hypersurface crossing the horizon under the same Max/Min conditions. 

It is first intriguing to  discuss one of the heuristic  arguments that is sometimes invoked in this context to conclude that the null and spacelike counting should be expected to yield the same result.

This argument generally goes as follows, \cite {Dou:2003af, Machet:2020uml}. 

Consider a one-parameter family of spacelike hypersurfaces 
$\Sigma_t$ which continuously deform to a null hypersurface,
$\Sigma =\lim_{t\rightarrow \infty} \Sigma_t$. 
On other hand, the region one sprinkles into and so the probability measure is also continuous with respect to the deformations. Now, because all spacelike hypersurfaces give the same result then the null hypersurface $\Sigma$ which can be casted  as a limit of a sequence of spacelike hypersurface should also give the same result. In the flat case one can equally use Lorentz invariance  of the counting and the fact any  spacelike hypersurface must give the same result, as any other related to it by a boost, and in the limit of tilting, a spacelike line becomes null. Note that a similar argument can also be made in the Schwarzschild case, using time-translation Killing vector instead of the boost killing vector.

  Stated mathematically, one would expect the following limit to hold
  
  \be\label{limitnullspace}
  \lim_{t\rightarrow \infty} 	<\mathbf{H}(\Sigma_t)>=<\mathbf{H}(\Sigma)> \ .
  \ee

  In the above equation we do not of course require strict equality, but it would be enough to hold in the limit $l_c\rightarrow 0$ modulo some statistical deviations from the leading mean value.  In \cite{Machet:2020uml}, it was for instance argued that it is the non-commutativity of the two limits, $ t\rightarrow \infty$ and $l_c\rightarrow 0$, which causes the above identity to fail, namely

  \be\label{limitnullspace2}
 \lim_{l_c\rightarrow 0} \lim_{t\rightarrow \infty} l_c^{d-2}<\mathbf{H}(\Sigma_t)>\neq \lim_{t\rightarrow \infty} \lim_{l_c\rightarrow 0} l_c^{d-2}<\mathbf{H}(\Sigma_t)> \ .
  \ee

   However, we shall see in the fourth section that in some horizon molecules counting the limit  $l_c \rightarrow 0$ is not at all required for the derivation of the area law, nonetheless the null and spacelike hypersurfaces give two different results. Therefore, we conclude that the above heuristic  argument invoking the non-ommutativity of the two limits is at best not generally sustainable,  as some counting could be inherently discontinuous and depend on the nature of the hypersurface crossing the horizon.
 
 	 \begin{figure}
 	 	\begin{center}
 	 		\includegraphics[height=4.1in,width=7in,angle=0]{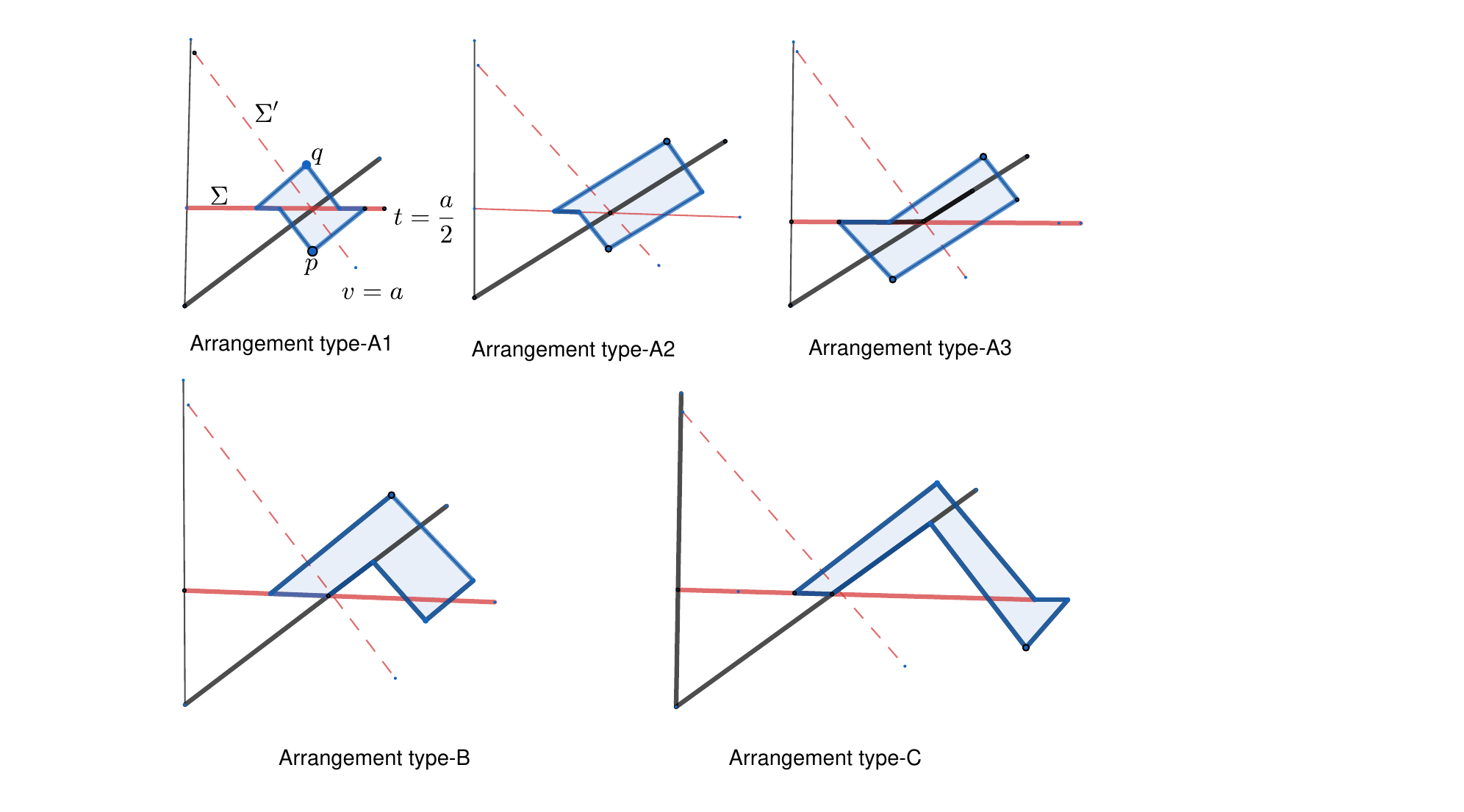}
 	 		\caption{Different arrangement contributing in the spacelike case, accorrding to different volume epxressions. For the arrangement type-A1 the point $q$ does not necessary lie to the future of $\Sigma '$, it could be to the past of it as well .}
 	 	\end{center}
 	 \end{figure}
 	\hfill \break
 Let us now consider a spacelike $\Sigma$  given by $t=\frac{a}{2}$. To facilitate  the discussion it is convenient to introduce a null hypersurface $\Sigma '$ defined by $v=a$. Again, we shall push the collapse to future infinity and restrict the counting to the flat region, Figure 5.
 	
 It is not difficult to see that one has to distinguish five cases, each having a different expressions for the volume $V(p,q)$. These cases are depicted in  Figure 5.

 The contributions $A1, A2, A3$ can be easily seen to be qualitatively of the same order of  magnitude as the null contribution we already evaluated, thus they  will just give  constants of order one, but surely  each is less then $\frac{\pi^2}{6}$.
 
 Contributions of type $(B)$ and $(C)$  are different and one can not directly conclude that are finite or of the order one. For instance contributions from pairs $(p,q)$  with $u_q\rightarrow 0$ and $t_p \rightarrow a/2$ could lead to IR divergences. However, it was  explicitly shown in \cite{Dou:1999fw} that both contributions are finite and of order one.
 
 Now, although it was possible to show that the expected number of such horizon molecules   give a constant of order one, the question whether the spacelike and null cases yield the same result  has so far remained open due to the analytical intractability of the integrals involved . But what  should be noted in this context is that  the difficulty to settle this issue in two dimensions  may not solely be of technical nature, due to the  intractability of the integrals, but there could be an other  issue of conceptual nature at work here. For example in the non-equilibrium case and for a null hypersurface counting, we argued that contributions coming from the pairs $(p,q)$ with $v_p<0$ are to be viewed as random fluctuations around the mean value not genuinely associated with $\Sigma \cap \mathcal{H}$   not to be included in the counting, despite the fact that they are not zero, but in higher dimension they are easily seen to be negligible for a macroscopic horizon. For the same reasons we expect similar  fluctuations to be present in  the above different contributions; in particular we do not expect arrangement of type-$(B)$ and $(C)$ to be fully associated with $\Sigma \cap \mathcal{H}$.  However, as it will be shown in the next section, the present links counting  fails to work beyond two dimensions due to IR divergences, and therefore pursuing this issue would be no more than mathematical curiosity without physical guidance nor relevance.

 		\subsection{The failure of the causal links proposal in higher dimensions }
 		
 	In view of the success and promising results of the causal links counting in two dimensional models, the natural step would  of course be to try to apply it on a more realistic four dimensional black hole background.  Any  direct attempt to do this 
 		counting  in Schwarzschild geometry   will inevitably be encountered by mathematical complications that are almost impossible to surpass. However, the results obtained previously in two dimensions enable us to transform the whole problem into calculation in  4- dimensional  (or $d$-dimensional ) flat spacetimes using the collapsing null-shell model by pushing the collapse world-line to future infinity. Therefore, one is in principle entitled to consider a flat spacetime with the future light-cone of the origin being the horizon, take a hypersuface $\Sigma$ (spacelike or null) intersecting the horizon and compute the expected number of horizon molecules, Figure 6 . 
 		
 	Although working in flat spacetimes drastically simplify the problem, in $d>2$   the calculation of $<\mathbf{H}_{link}>$ is still complicated enough and a much more elaborated
 		technique is needed to do the counting explicitly,  for both the null and spacelike hypersurfaces. The calculation of the
 		volumes needed to insure the link and Max/Min conditions  is lengthy; and it turns out that
 		one has to distinguish many cases depending on the relative positions of the points $p$ and $q$, each case making its own contribution to $<\mathbf{H}_{link}>$ \cite{Dou:1999fw}.  For instance, for  spacelike $\Sigma$  and to the exception of one contribution, coming from an  arrangement similar to type A.1 in two dimensions, Figure 5,  which  could be evaluated and was reported in \cite{Dou:1999fw, Dou:2003af};  the remaining arrangements turned out to be either very complicated or intractable. Nonetheless, at least for one  non-trivial arrangement the volume was computed exactly in \cite{Dou:1999fw}. The  arrangement in question is of the type-B , depicted in Figure 5 (its four dimensional analogue) . For this particular arrangement it was later realized by the author  that its  corresponding contribution to    $<\mathbf{H}_{link}>$ diverges . It is unnecessary to give the detail of  this calculation, but it is not difficult to qualitatively  understand the source of this IR divergence .
 		\begin{figure}
 			\begin{center}
 				\includegraphics[height=4.1in,width=6.8in,angle=0]{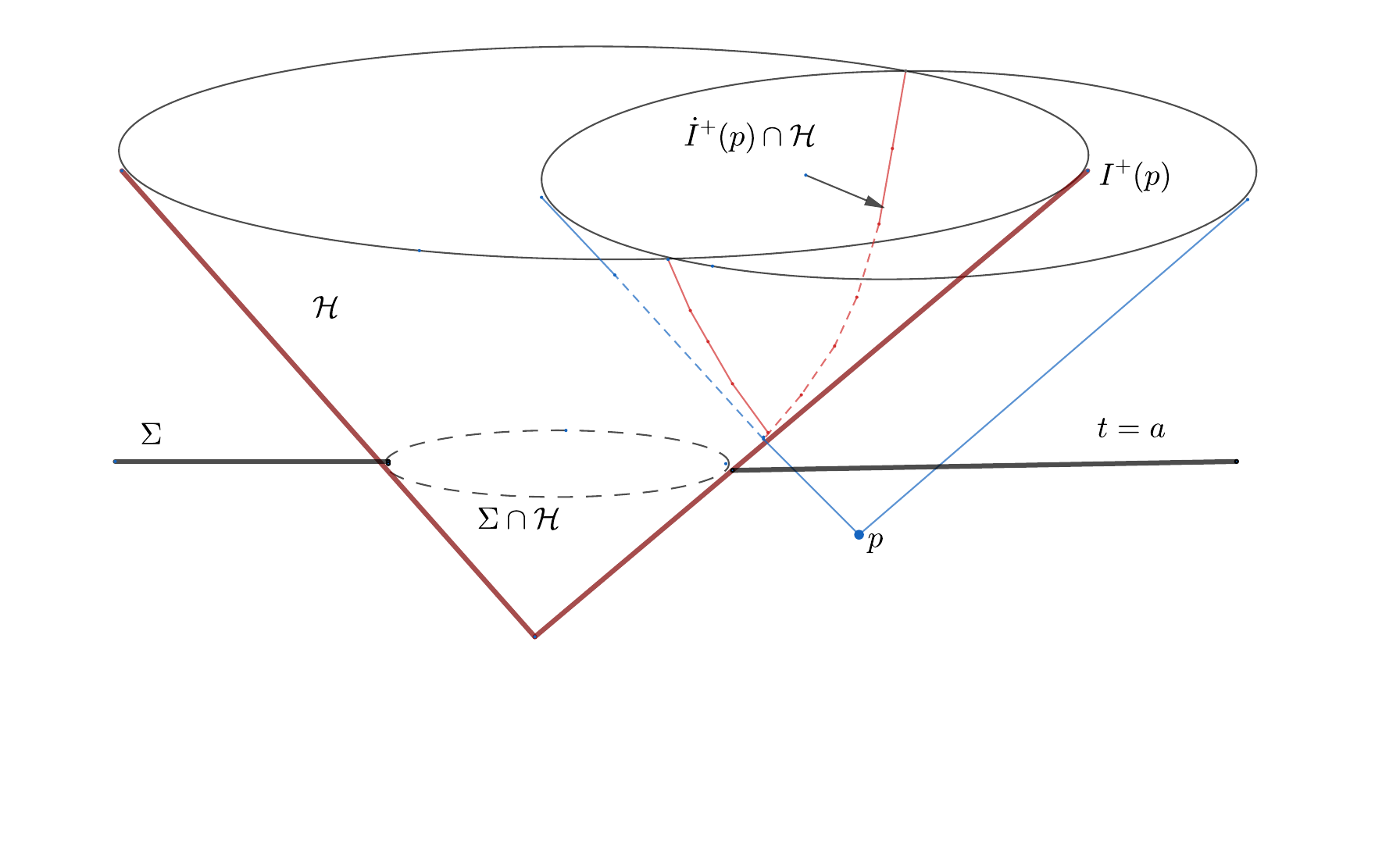}
 				\caption{The red curve shows the intersection of $\dot{I}^+(p)$ with the horizon, this curve extends to future infinity and by considering sprinkled points $q$  asymptotically approaching this curve, an arbitrarily number of $p\prec \!\!\cdot  \, q$ links can be found.}
 			\end{center}
 		\end{figure}
 		
 		Let us first take a step back and consider the arrangement Type-B depicted in Figure 5, in its 1+1 version. 
 		The only potential source of IR divergences comes from the contributions of  points $q$ arbitrarily far away from the intersection point  $\Sigma\cap\mathcal{H}$; however the  $e^{-V}$ term  appearing in the integrand exponentially suppresses all contributions except those when $p$ is arbitrarily close to the intersection point and $q$ bound to the horizon, which is not enough to produce any IR divergences. 
 		
 		The situation in higher dimensions is quite different and this can be grasped by considering the 2+1 case depicted in Figure 5,  and using the qualitative argument  given in \cite{Sarah}.

 		In Figure 6  the horizon light cone $\mathcal{H}$ is intersected by  a spacelike hypersurface $\Sigma$, $t=a$ say. Consider a point $p$ and let $\dot{I}^+(p) $ be the boundary of its  future light-cone . Unlike 1+1, in 2+1 dimensions (or higher)  the intersection of $\mathcal{H}$ is no longer a point, but rather a curve ( $d-2$ dimensional surface in general). This adds a new degree of freedom and allows the existence of new  links formed with points $q$, asymptotically close to $\dot{I}^+(p)\cap \mathcal{H} $ and arbitrarily far form $\Sigma\cap \mathcal{H}$. In addition, and unlike the 1+1 case, the point $p$ is not at all required to be arbitrarily  close to the $\Sigma\cap \mathcal{H}$  for the volume $V$ to vanish; it is enough to be arbitrarily close to $\Sigma$ and anywhere far from the intersection of $\mathcal{H}$ and $\Sigma$. For these  distant pairs $(p,q)$, the interval between $p$ and $q$ remains small and highly likely to be free from additional sprinkled points.   For these reasons the $e^{-V}$ is not enough to suppress the contributions of such pairs, as there is an infinite number of potential pairs $(p,q)$ with vanishing volume in the analytic limit.  As a consequence the expected number of causal links, no matter which  Max/Min conditions are imposed, will diverge like $\Lambda_r^{d-2}$, in $d$-dimensions, with $\Lambda_r$ an appropriate IR cutoff. This IR divergence is incurable within the causal links proposal and a real departure from the links definition  is therefore inevitable .

 		\section{The triplet proposal}
 		
 		The failure of the causal links proposal  beyond 1+1  subsequently led soon to a different and modified causet structures as  a new candidates for the horizon molecules. 
 		
 		We first note  that  it is almost obvious that the day can not be saved  by  simple modifications of the links proposal, for instance by modifying the Max/Min conditions, as we have exhausted all acceptable variants. The first attempt to depart   from the links structure   was  taken by Marr \cite{Sarah}. This attempt was mainly based on "triad" structure or "Triplet". Although there are some hints that this modified horizon molecular structure may fail to cure all the IR divergences we observed in the causal links proposal  in higher dimensions,  we find that the triplet proposal  worthy of brief  discussion.  We shall omit all technical details, as it is similar in spirit to the link-counting, and only focus on the main results and their discussion.

 		Let us start by noting that there actually some suggestion that 
 		a certain type of triplets are naturally related to the kind of correlation responsible for entanglement entropy in a quantum field theory framework \cite{Sorkin:1995nj}.

 		\textbf{$\Lambda$-Triplet (Marr 2007)}:

 		A horizon molecule with respect to  a given hypersurface $\Sigma$ is  a triplet $(p,q,r)$ satisfying the following conditions
 		
 		\begin{enumerate}
 			\item $p \in I^-(\Sigma ) \cap I^-(\mathcal{H})$ ,
 			\item $q \in I^+(\Sigma ) \cap I^+(\mathcal{H})$,
 			\item $r \in I^-(\Sigma ) \cap I^+(\mathcal{H})$,
 			\item $|I[p,q]|=0$, $|I[r,q]|=0$ .
 			\item $p$ is maximal in $I^-(\Sigma ) \cap I^-(\mathcal{H})$ , $q$ is minimal-but-one in $I^+(\mathcal{H})$ and $r$ is minimal in  $I^-(\Sigma ) \cap I^+(\mathcal{H})$  .
 		\end{enumerate}

 		Notice that the $4^{th}$ condition automatically requires $p$ and $r$ to be causally unrelated  or spacelike related. 
 		
 The condition  $q$ minimal-but-one in $I^+(\mathcal{H})$ is meant to ensure that the only point in $I^-(q )$ is $r$, Figure 7.

\begin{figure}
	\begin{center}
		\includegraphics[height=3.3in,width=6in,angle=0]{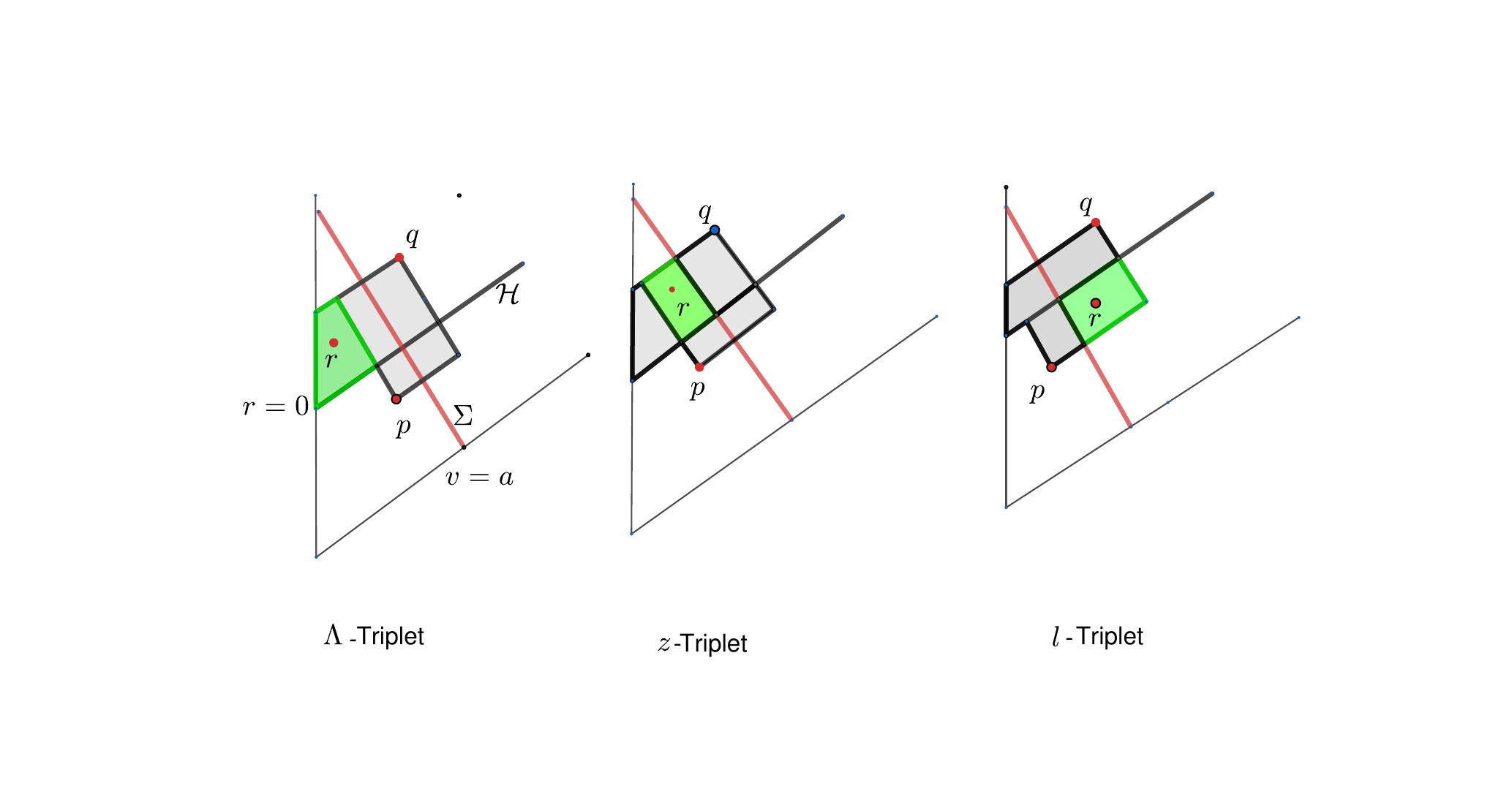}
		\caption{The  regions shaded grey are required to be free from any sprinkled points, whilst the green ones have only one point .}
	\end{center}
\end{figure}

 Marr  first applied the $\Lambda$-triplet proposal for the collapsing shell model in 1+1 - pushing the collapsing shell to future infinity- to obtain an expected number of horizon molecules  of order one, more precisely she obtained
 
 \be\label{L-T}
 <\mathbf{H}_{\Lambda-triplet}> = {\pi^2 \over 6}-1 + \mathcal{O} ((a/b)^2)  \ .
 \ee

 Using a triplet instead of a simple link (doublet) has  therefore reduced the expected number of horizon molecules by one.
 
 In contrast to the collapsing shell model in $1+1$, the case of $2$-d reduced Schwarzschild BH turned out to be analytically intractable for the $\Lambda$-triplet.
 
As  mentioned earlier, the underlying motivation that led to the departure from the link structure to the triplets was to kill off the contributions coming from $p-q$ links generated by points $q$ arbitrarily close to $\mathcal{H}\cap \dot{I}(p)$ but arbitrarily  far from $\mathcal{H}\cap \Sigma$. However, it has been argued in \cite{Sarah} that although the introduction of a third element in the horizon molecule structure, i.e. $r$, seems to cure this IR divergences,  the $\Lambda$-triplet counting still suffers from another IR divergences, of course already present in the links proposal. This  IR divergence arises   
 as a result of having unsuppressed contributions coming now from points $p$ asymptotically approaching $\Sigma\cap \dot{I}(q)$  as they move further into the past, simultaneously keeping $p-q$ interval small, and as $r$ does not bound $p$ from $\Sigma$, they are spacelike related, thus avoiding any exponential suppression .
 
 The above qualitative argument seemingly  rules out the $\Lambda$-triplet as possible alternative candidate that would work in higher dimensions, and led Marr to consider other possible arrangements  for the triplet. The guide of course  was to cure the IR divergences which plagued the link-counting and persisted in the $\Lambda$-triplet  counting. 
 
 To that end two different arrangements were considered in \cite{Sarah},  the $z$ and $l$-triplet.

  The $z$-triplet is  obtained from the definition of the $\Lambda$-triplet by keeping the first three conditions, moving $r$ to the future of $p$ to  form a link with it,  keeping its link relation with $q$, so the three points form a path or a maximal chain, i.e $ p\prec \!\!\cdot  \,r\prec \!\!\cdot  \,q$, and removing  the minimality condition on $r$ from the $5^{th}$ condition, Figure 7.
  
  As for the $l$-triplet, it is a re-arrangement of the $z$-triplet by moving $r$ to region $ \in I^+(\Sigma ) \cap I^-(\mathcal{H})$, requiring $p$ to be maximal in $I^-(\Sigma ) \cap I^-(\mathcal{H})$ and $q$ maximal in $I^+(\mathcal{H})$, Figure 7.
  
  Marr used both the $l$ and $z$-triplet to count the expected number of horizon molecules for the collapsing null shell 1+1 reduced model and obtained the following results
  
  \be
   <\mathbf{H}_{z-triplet}> = 2-{\pi^2 \over 6} + \mathcal{O} ((a/b)^2)  \ ,
  \ee
 	\be
 	<\mathbf{H}_{l-triplet}> = 1+\mathcal{O} (a/b) \ .
 	\ee

 	Moreover, the $l$-triplet turned out to be manageable 
 	analytically for $1+1 $  Schwarzschild static model and gave the same result (leading term) as the collapsing null shell setting, which is a promising result.
 	
 	Let us remember that both the $z$ and $l$-triplet were introduced with an eye on their use as a candidate for horizon molecules in higher dimensions. Although Marr did not report any analytical  results  concerning the  triplet structures in 1+2 or 1+3, she gave  qualitative argument suggesting  that the $z$ and $l$-triplet are in principle free of the IR divergences we discussed above. Her argument goes as follows: The presence of a third element $r$ in the $z$-triplet bounds $p$ away from $\Sigma$  and $q$ from $\mathcal{H}$, whereas in the $l$-triplet the role of $r$ is reversed, it bounds $p$ away from $\mathcal{H}$    and $q$ from  $\Sigma$. Hence for both triplets an arbitrarily large number of $p-q$ links is unlikely to build up in higher dimensions .  
 	
 	Let us note that Marr's qualitative  argument regarding the would-be role  played by $r$ in killing off the IR divergences in higher dimensions does not guarantee  the finiteness of $ <\mathbf{H}>$ nor the emergence of the area law. For there could exist other less obvious and more subtle sources of divergences.  Moreover, the finiteness of the result does not either guarantee that the resulting $ <\mathbf{H}>$ will scale like the area.  Therefore the matter can only be 
 	settled by explicit calculation and this brings in a technical difficulties that one has to deal with when considering higher dimensions, and in particular 3+1. These technical difficulties come from the necessity to evaluate the volumes needed to insure links and max/min conditions, which turned out to be complicated and in some cases intractable even in the flat case and for the link structure \cite{Dou:1999fw}, let alone the triplet structure . However, there are indications that the case $2+1$ could be easier to handle analytically. Thus it would be interesting to explicitly test the triplets proposals, in particular the $z$-triplet in $2+1$ is worth revisiting . 
 	
 	It    is also worth mentioning  that $z$ or $l$-triplets could work in $2+1$ but fail beyond this dimension, and one may be led to consider ``diamond" structure containing both types of triplet simultaneously.

\section{An extended notion of horizon molecules}

The definition of the horizon molecules as simple causal links crossing the horizon, supplemented with certain Max/Min conditions, worked nicely and gave promising results in $2$-d reduced spacetimes, at least for null hypersurfaces. However, this success did not carry over to higher dimensions due to pathological IR divergences.  The higher cardinalilty molecule definitions, namely the triplets,  proposed by Marr  to cure these divergences turned out to be mathematically cumbersome and challenging beyond two dimensions, and so far no one has devised a  technique that would allow analytical investigation of the triplets proposal in higher dimensions. This  calculational  impasse, the failure of the causal links proposal  and the desire
to extend the concept of horizon molecule to all causal horizons including black
hole, acceleration, and cosmological horizons stimulated  two recent sequential and related works by Barton et al  \cite{Barton:2019okw} and by Machet and Wang \cite{Machet:2020uml}. These two works  will be the subject of the present section.

 Our discussion will not cover the  technical details presented in \cite{Barton:2019okw, Machet:2020uml}, but will be limited to introducing the key technical ideas, the results obtained and their discussion.

\subsection{The spacelike hypersurface case}
 	 
 	 The  extended proposal put forward by Barton et al  was  devised to extend the notion of horizon molecule to more general causal horizons, thus it does not refer to any particular black hole geometry, and to produce the area law in higher dimensions.  
 	 
 	 The definition goes as follows.   Let $(\mathcal{M}, g)$ be a globally hyperbolic spacetime with a Cauchy
 	 surface $\Sigma$. Let $\mathcal{H}$ be a causal horizon, defined as the boundary of the past of a future inextendible timelike
 	 curve $\gamma$, i.e.$ \mathcal{H} := \dot {I}^-(\gamma)$, and consider a causet $\mathcal{C}$  generated on $\mathcal{M}$ through a random sprinkling
 	 sprinkling with a density $\varrho_c$. 
 	 
 	 \hfill \break
 	 \textbf{Barton et al proposal (2019)  }
 	 A horizon molecule with respect to space-like hypersurface $\Sigma$  ( a Cauchy surface) is a pair of elements of $\mathcal{C}$, $\{p_-,p_+\} $, such that
 	
 	\begin{enumerate}
 		\item  $p_-\prec p_+$  ,
 		 \item $p_-\in I^-(\Sigma ) \cap I^-(\mathcal{H})$  ,
 		\item $p_+ \in I^-(\Sigma ) \cap I^+(\mathcal{H})$ \ ,
 		\item $p_+$ is the only element in  $I^-(\Sigma )\cap I^+(p_-)$  .  
 	\end{enumerate}
 
 Theses conditions imply that the horizon molecule is a link. An illustration of this type of horizon molecules is depicted in Figure 8.
 
\begin{figure}
	\begin{center}
		\includegraphics[height=3.3in,width=5in,angle=0]{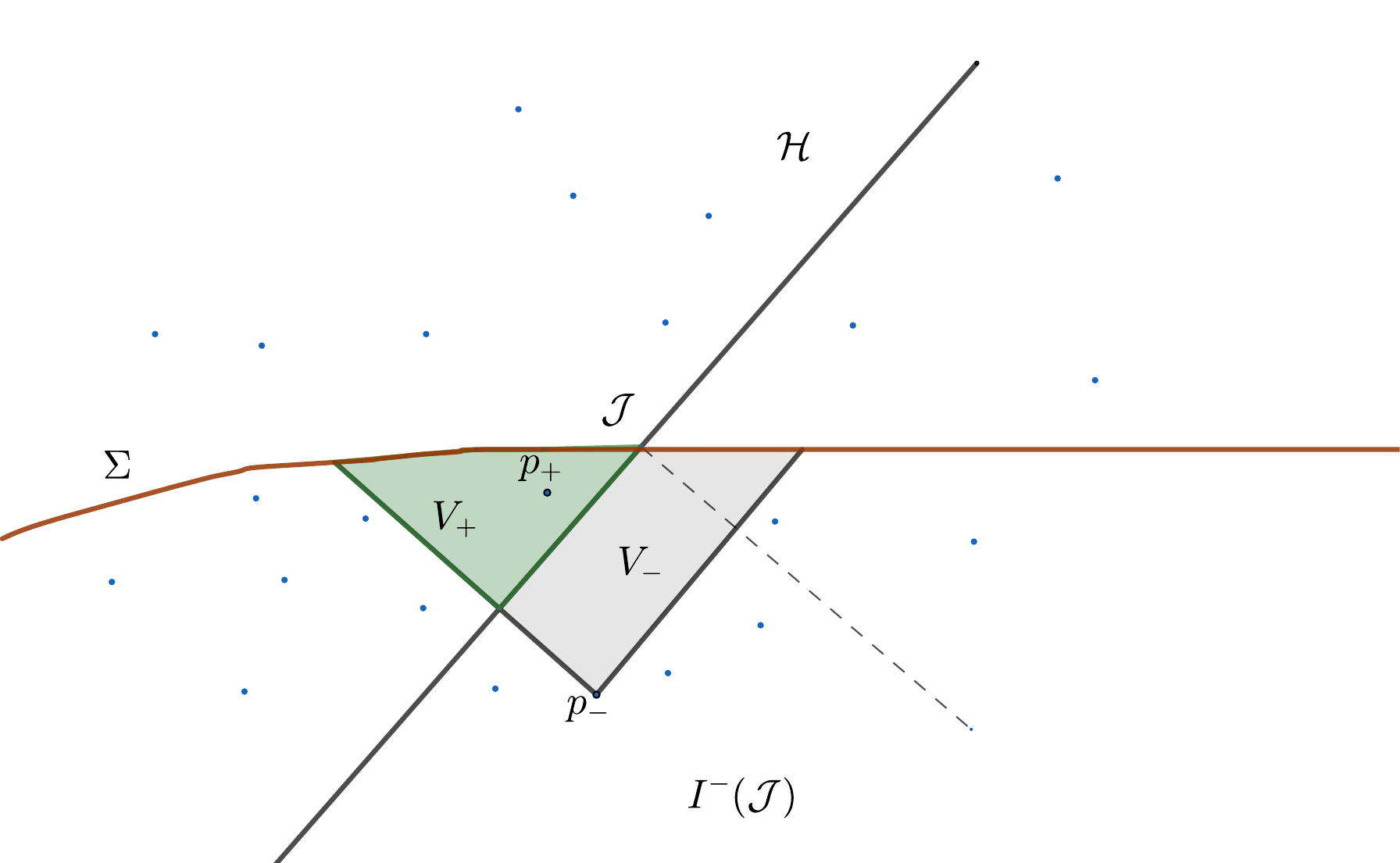}
		\caption{A typical geometrical setting showing a typical horizon molecule $(p_-,p_+)$ ($n=1$).}
	\end{center}
\end{figure}

 The above definition is easily seen to be  generalizable to $n$-molecule $\{p_-,p_{1,+}, p_{2,+},\cdot \cdot \cdot ,p_{n,+}\} $ by requiring $p_-\prec p_{+,k}$ and $\{ p_{1,+}, p_{2,+},\cdot \cdot \cdot ,p_{n,+}\} $ to be the only elements in  $I^-(\Sigma )\cap I^+(p_-)$ . However, our discussion of this proposal will be limited to the horizon molecule of minimal size $n=1$. Actually the cardinality of the horizon molecules plays no essential technical role in the derivation of the results obtained in \cite{Barton:2019okw}.

 Before we move to the discussion of the derivation of the results of Barton et al,  we find it instructive to compare the above definition  with the original definition of horizon molecules as causal links with certain Max/Min conditions.
 
 The extended horizon definition  requires $p_-$ to be a maximal element in $I^-(\Sigma ) \cap I^-(\mathcal{H})$, or maximal-but-one in $I^-(\Sigma)$, and no similar maximality condition is imposed on $p_+$.  The requirement that $p_-$ be maximal-but-one (or but-$n$) would for instance derive the expected number of horizon molecules directly to zero if the hypersufrace was null (straight null plane) and the future of the horizon is unbounded, as it is the case for Rindler space for example. For this  and others reasons the null case has motivated an independent  work  by Machet and Wang \cite{Machet:2020uml}, to which we will come lastly.
 
 %Stated differently, the expected number of horizon  molecules   to be counted here is the expected number of points   $p_-\in I^-(\Sigma ) \cap I^-(\mathcal{H})$ which are maximal-but-one in $I^-(\Sigma )$.

 It  was first shown in \cite{Barton:2019okw} that $p_-$ is in the chronological past of $\mathcal{J}=\Sigma \cap \mathcal{H}$. Using similar steps we  used to arrive to (\ref{HMF}),  it is not difficult to obtain the following integral representation for the expected number of such horizon molecules,
 
 \be\label{Hextended}
 <\mathbf{H}_1>=\varrho_c \int_{I^-(\mathcal{J})} \rho_c  V_+(p) e^{-\varrho_c V(p)} dV_p \ ,
 \ee
 
 where  
 $$
 V_+(p):= \text{vol}(I^-(\Sigma ) \cap I^+(\mathcal{H})\cap I^+(p)),~~V_+(p):= \text{vol}(I^-(\Sigma ) \cap I^+(p)) \ ,
 $$
 
 and  $p$ is used to denote the point $p_+$. Figure 8 illustrates the   volumes involved in the counting.

 The main result shown by Barton et al is that under certain conceivable assumptions and in the continuum limit, the expected number of such defined horizon molecules, suitably rescaled, is equal to the area of  $\mathcal{J}$, the intersection of $\Sigma$ and $\mathcal{H}$, up to a dimension dependent constant of order one. Mathematically stated  we have the following limit

 \be\label{expHext}
 \lim_{\varrho_c\rightarrow \infty} \varrho_c^{\frac{2-d}{d}}<\mathbf{H}_1>= a^{(d)}\int_{\mathcal{J}} dV_{\mathcal{J}} \ ,
 \ee

 where $dV_{\mathcal{J}}$ is the area measure on $\mathcal{J}$, and $ a^{(d)}$ is a constant that only depends on the dimension $d$.  Moreover, the approach to the limit involves finite $\varrho_c$ corrections forming  a derivative expansion of local geometric quantities on $\mathcal{J}$ and increasing powers of $l_c$, the discreteness length. In  what follows we shall therefore explicitly keep track of $\varrho_c$ and $l_c$.

\subsubsection{Rindler Horizon with a flat hypersurface} 
 Before outlining the explicit calculation of \cite{Barton:2019okw}, it would be instructive to present their heuristic argument supporting the validity of  (\ref{expHext})  in general.  This heuristic argument actually summarizes the motivation behind defining the horizon molecules as such.
 
 Consider Figure 8, the fact that $p_-$  lies in $I^-(\mathcal{J})$ and is required  to be
 maximal in this region means that $p_-$ is close to $\mathcal{H}$, and as $\varrho_c \rightarrow \infty$ it gets closer. The requirement that $p_-$ is maximal-but-one in $I^-(\Sigma ) \cap I^+(\mathcal{H})$ pushes $p_-$  towards $\Sigma$ and prevents it from moving  to the past of $\mathcal{J}$.

  This tendency can be seen by inspecting the integrand of (\ref{expHext}) in which the  exponential  will suppress any contribution from regions with $\varrho_c V(p)\gg 1$. Contributions coming from points far from the horizon are suppressed by the $e^{-\varrho_c V_-(p)}$, whereas those close to the horizon but far from $\Sigma$ are suppressed by $e^{-\varrho_cV_+(p)}$. Therefore the only region which gives a non-negligible contribution is a small and decreasing subregion of $I^-(\mathcal{J})$, immediately to the past of $\mathcal{J}$. This strongly  suggests that in the limit, the integral will only depend on geometric quantities intrinsic to $\mathcal{J}$. On dimensional ground, the only  geometric quantity that can appear on the RHS of (\ref{expHext})  is the area of  $\mathcal{J}$ times a dimensionless constant, $a^{(d)}$, which is independent of 	the geometry.

 		To prove (\ref{expHext}) Barton et al first proceeded by probing their defintion on Minkowski $d$-dimnsional spacetime,  flat $\Sigma$ and Rindler horizon $\mathcal{H}$, in short all-flat.
 		
 		First  an inertial coordinates system   is set up, $(x^0, x^1, y^\alpha), \a =2,3,\cdots d-1$ , the hypersurface $\Sigma$ is chosen at $x^0=0$. $ \mathcal{H}$ is given by $x^0=-x^1$. For technical convenience   a past-futue sawpped set up was instead used, so the domain of integration is $p\in I^+(\mathcal{J})$.  The integrand is independent of $y^\alpha$ and the scaled expected number of horizon molecules takes the form
 		
 		\be
 		 \varrho_c^{\frac{2-d}{d}}<\mathbf{H}_1>= \int_{\mathcal{J}} d^{d-2}y I^{d,flat}(l_c) \ ,
 		\ee

 		where $I^{d,flat}(l_c)$ is a dimensionless function given by
 		
 		\be\label{I1}
 		I^{(d,flat)}(l_c)= l_c^{-(d+2)} \int_{0}^{\infty} dx^0 \int_{-x^0}^{x^0} dx^1 \tilde{V}_+(x) e^{-\varrho_c \tilde{V}(x)} \ ,
 		\ee

 	 where $l_c=\varrho_c^{-1/d}$ is the discreteness length.  In the limit $l_c\rightarrow 0$, the function $	I^{(d,flat)}(l_c)$ determines the constant $a^{(d)}$.

 	 In this flat case $\tilde{V}(x)$  is just the $d$-dimensional volume of a solid null 
 	 cone of height $x^0$. 
 	 For $\tilde{V}_+(x)$,  it was not possible to derive a formula for general
 	  dimension $d$, but it was possible to compute it explicitly for the lower dimensions, $d=2,3$ and $ 4$, and they are given respectively by
 	  
 	  \bea
 	   d=2 &:& \tilde{V}_+(x)=\frac{1}{4}(x^0-x^1)^2 \ , \nonumber\\
 	 d=3 &:& \tilde{V}_+(x) = \frac{2}{3}(x^0)^3 \tan^{-1} \big( \sqrt{\frac{x^0-x^1}{x^0+x^1}}\big) \nonumber\\
 	 &-& \frac{1}{9}(2x^0-x^1)(2x^0+x^1) \sqrt{(x^0-x^1)(x^0+x^1)} \ , \nonumber\\
 	 d=4 &:&  \tilde{V}_+(x)=\frac{\pi}{48}(x^0-x^1)^3(5x^0+3x^1) \ . \nonumber
 	 \eea

 	 A direct, but not straightforward, calculation leads to the following  limits
 	 
 	 $$
 	 \lim_{l_c\rightarrow 0} I^{(2,flat)}(l_c)=a^{(2)}=\frac{1}{3} \ ,
 	 $$
 	 
 	 $$
 	  \lim_{l_c\rightarrow 0} I^{(3,flat)}(l_c)=a^{(3)}=\frac{1}{4}\big( \frac{3}{\pi}\big)^{2/3} \ ,
 	  $$
 	  
 	  $$
 	  \lim_{l_c\rightarrow 0} I^{(4,flat)}(l_c)=a^{(4)}=\frac{\sqrt{3}}{10} \ .
 	  $$
 
 Actually the constants $a^{(d)}$ were given  in \cite{Barton:2019okw} for arbitrary $n$.

 It should be noted here that although Barton et al computed the constants $a^{(d)}$ in the limit $l_c\rightarrow 0$  using Watson's lemma, the function $	I^{(d,flat)}(l_c)$ is independent of the discreteness length $l_c$ and equals  $a^{(d)}$   at any discreteness scale . In other words we have
	\be\label{I12}
I^{(d,flat)}(l_c):=I^{(d,flat)}= a^{(d)}= l_c^{-(d+2)} \int_{0}^{\infty} dx^0 \int_{-x^0}^{x^0} dx^1 V_+(x) e^{-\varrho_c V(x)} \ ,
\ee
 
 and the above particular numerical values for $ a^{(d)}$ are the exact values of the integrals  $I^{(d,flat)}(l_c)$ for different $d$ regardless of the value of $l_c$.
 
There are indeed two ways to see why   $I^{(d,flat)}(l_c)$ must be independent of  density of the sprinkling $\varrho_c$ or $l_c$.  The first is purely technical and based on a simple dimensional analysis of the integral (\ref{I1}). As $I^{(d,flat)}(l_c)$ is dimensionless,  and there is no length scale which can pair with $l_c$ to form a dimensionless quantity, thus the result must be a pure number.

Another heuristic, but more intuitive, argument to understand  why the derivation of the above area law should be independent of  the density of the sprinkling in the all-flat setting, is the following. 

  In an all-flat setup,  the two regions $ I^-(\Sigma ) \cap I^-(\mathcal{H})$ and 
   $ I^-(\Sigma ) \cap I^+(\mathcal{H})$ are flat and infinite (unbounded). If one randomly sprinkles in points in both regions with a given density, $\varrho_1$ say, and  considers another sprinkling with density $\varrho_2$, both sprinklings should give the same result. The situation  is just a matter of zoom -in  and zoom -out, the trade-off  here is simply that  the number of molecules one loses by decreasing the density of sprinkling,  gains by moving further to the past ( away from the intersection of the horizon and $\Sigma$). The density  only tells us how far into the past we should go for  the value of  the constant $a^{(d)}$ to get effectively saturated,   and in the very large  density limit the molecules contributing to $<\mathbf{H}_1>$ are the ones located infinitesimally close to $\Sigma  \cap \mathcal{H}$   . In other words, if the integral over $x^0$ in (\ref{I1}) 
   were cut off at some upper limit $\tau \gg l_c$, the value of the $ l_c\rightarrow 0$ limit would not be affected, the deviation from the above limiting values tends to zero exponentially fast. This locality property, the fast exponential vanishing of the difference, will be crucial for the discussion of the general curvature case to which we now turn.

  \subsubsection{The general curvature case}
  Our discussion of the general curvature case will more or less be sketchy, escaping  technical detail and just highlighting the crucial steps of the calculation of Barton et al.
  
  The key elements in proving the limit (\ref{Hextended})  in the general curvature setting were first the construction  of Florides-Synge Normal Corrdinates (FSNC'S)  based on the co-dimension $2$ spacelike submanifold  $ \mathcal{J}$ and second the locality argument. Such coordinates system construction is always possible in tubular neighborhood about a submanifold of any co-dimension in any Riemannian or pseudo-Riemannian manifold \cite{doi:10.1098/rspa.1971.0085}.
  
For $d>2$,  let $z^a =(x^A, y^\alpha) (A=0,1, \a= 2,\cdots ,d-1)$ denote the FSNC's corrdinates  constructed within a small enough tubular neighborhood $ \mathcal{N}$ about $ \mathcal{J}$.  For $d=2$ FSBC's are just the Riemann Normal Corrdinates based on the intersection point $ \mathcal{J}$.

The next step is to assume the existence of a length scale $\tau$ such that $ l_c\ll \tau \ll L_G$, where $L_G$ is the smallest geometric scale in the setup.  This assumption is reasonable, because  the continuum approximation of causal set is only valid when the curvature length scales involved in the problem are much larger than the discreteness scale $l_c$.

Consider now the region $ \mathcal{R}_
\t$ defined as
 
$$
\mathcal{R}_\t:=\{ p\in I^-(\mathcal{J}) \cap \mathcal{N}~: -\t <x^0(p)<0\}\ ,
$$
  
 where $\t$ is assumed to be small enough that this region is inside the tubular neighborhood $\mathcal{N}$.
 
 Let $\bar{\mathcal{R}}_
 	\t := I^-(\mathcal{N}) \setminus \mathcal{R}_\t $ denote the complement of 
$\mathcal{R}_\t$, then the integral (\ref{expHext}) naturally splits into a part over $\mathcal{R}_
 	\t$, and another over $ \bar{\mathcal{R}}_
 	\t$.
 
 Now, in \cite{Barton:2019okw} it was argued, using the locality argument, that the integral over  $ \bar{\mathcal{R}}_\t$  tends to zero faster than any power of $l_c$, actually exponentially suppressed,  and hence its contribution can be ignored.   Therefore the surviving part of the expected value can be written as a local integral over  $\mathcal{R}_\t$ 
 
 \be\label{Hextended2}
 \varrho_c^{\frac{2-d}{d}}<\mathbf{H}_1>=\varrho_c^{\frac{2-d}{d}+2} \int_{\mathcal{R}_\t}   V_+(p) e^{-\varrho_c V(p)} dV_p \ .
 \ee

In view of the fact that the region  $\mathcal{R}_\t $ lies by choice within the tubular neighbourhood, and hence the constructed FSNC's  can be used to express the expectation value explicitly  as

\be\label{Hextended2}
\varrho_c^{\frac{2-d}{d}}<\mathbf{H}_1>=\varrho_c^{\frac{2-d}{d}+2} \int_{\mathcal{J}} d^{d-2}y \int_{-\t}^{0} dx^0 \int_{x^0}^{-x^0} dx^1 \sqrt{-g(x,y)} V_+(x,y) e^{-\varrho_c V(x,y)} \ ,
\ee

where $g(x,y)$ is the determinant of the metric.

Let $\sigma_{\alpha \beta}$ denote  the induced metric on $\mathcal{J} $, then (\ref{Hextended2})  can be written as 

\be\label{Hextended3}
\varrho_c^{\frac{2-d}{d}}<\mathbf{H}_1>= \int_{\mathcal{J}} d^{d-2}y \sqrt{-\sigma(y)} I^{(d)}(y;l_c,\t)+\cdots \ ,
\ee

where $ I^{(d)}(y;l_c,\t)$ is defined by

\be\label{I2}
I^{(d)}(y;l_c,\t):=l_c^{-(d+2)}  \int_{-\t}^{0} dx^0 \int_{x^0}^{-x^0} dx^1 \sqrt{\frac{-g(x,y)}{\sigma(y)}} V_+(x,y) e^{-\varrho_c V(x,y)} \ ,
\ee
where $\sigma(y)$ is the determinant of the induced metric. 

The factor $\sqrt{\frac{-g(x,y)}{\sigma(y)}}$ makes $ I^{(d)}(y;l_c,\t)$ a scalar on  $\mathcal{J} $ and can be rewitten in a free coordinate notation as $ I^{(d)}(q;l_c,\t)$, $q\in \mathcal{J}$.
 
The next crucial step in the calculation of Barton et al is to show that $ I^{(d)}(q;l_c,\t)$  admits the following local expansion
\be\label{Iexp}
 I^{(d)}(q;l_c,\t)=a^{(d)}+l_c \sum_i b_i^{(d)} \mathcal{G}_i(q) + O(l_c^2) \ ,
 \ee
 
  where $a^{(d)}$ and $b_i^{(d)}$ are  constants that only depend upon the dimension $d$. For instance $a^{(d)}$
 	 is the same constant obtained in the flat case. $\mathcal{G}_i(q)$ is the largest set of mutually independent geometric scalars of length dimension $L^{-1}$, like the extrinsic curvature $K$  or the null expansion $\theta$ evaluated at $q$. 
 	
 	Again,  switching  to an order-reversed  setup  $I^{(d)}(q;l_c,\t)$   is written  as
 	
 	 \be\label{I3}
 	 I^{(d)}(y;l_c,\t)=l_c^{-(d+2)}  \int_{0}^{\t} dx^0 \int_{-x^0}^{x^0} dx^1 \sqrt{\frac{-g(x,y)}{\sigma(y)}} V_+(x,y) e^{-\varrho_c V(x,y)} \ .
 	 \ee

 At  this stage one is free to choose any  coordinates on $\mathcal{J}$, and a suitable choice is RNC's  $y^\a$  centered about $q\in \mathcal{J}$; $y^\a(a)=0$. As all expressions  appearing in (\ref{I3}) are evaluated at $y^\a(a)=0$, the  argument $y$ will be dropped entirely to write 
   
 \be\label{I4}
 I^{(d)}(q;l_c,\t)=l_c^{-(d+2)}  \int_{0}^{\t} dx^0 \int_{-x^0}^{x^0} dx^1 \sqrt{-g(x)} V_+(x) e^{-\varrho_c V(x)} \ .
 \ee

  Note that $\sigma(0) =1$ in these RNC's on $\mathcal{J}$, $\sigma_{\a\b}(0)=\delta_{\a\b}$.
  
   Now spacetime  RNC's  $Z^{a}=(X^A, Y^\a) $ can  be introduced within a neighbourhood $\mathcal{U}$ about $q$ , such that $X^A=x^A$, and such that the coordinate vectors $\frac{\partial }{\partial Y^\a  } =\frac{\partial }{\partial y^\a  }$ at $q$. With this choice the determinant of the metric, evaluated at at $q$ keeps the same form in terms of the coordinates $x^A$ and $X^A$; and we have
   \be\label{I5}
   I^{(d)}(q;l_c,\t)=l_c^{-(d+2)}  \int_{0}^{\t} dX^0 \int_{-X^0}^{X^0} dX^1 \sqrt{-g(X)} V_+(X) e^{-\varrho_c V(X)} \ .
   \ee

     The determinant $g(X)$ can be expanded in small $X^A$ relative to the curvature scales of spacetime at $q$:
     \be
     \sqrt{=g(X)} =1-\frac{1}{6} R_{AB}X^AX^B+O(Z^3) \ ,
     \ee
     where $R_{AB}$ is the Ricci tensor with indices restricted to $A,B=0,1$. To bring out the role of the different length scales of the problem, the smallest length scale $L_G$ is used  to define a dimensionless tensor $\hat{R}_{ab}:=L_G^2R_{ab}$, and $\t$ is used to re-express the above expansion in terms of dimensionless coordinates $\hat{Z}_a:=Z^a/\t$  
     
 	 \begin{eqnarray}\label{metrixexp}
 	 	\sqrt{-g(X)} &=&1-\frac{1}{6} (\frac{\t}{L_G})^2\hat{R}_{AB}\hat{X}^A\hat{X}^B+O(Z^3)\nonumber\\
 	 	&=&1-\frac{1}{6} \varepsilon^2\hat{R}_{AB}\hat{X}^A\hat{X}^B+O(\varepsilon^3) \ .
 	 \end{eqnarray}

 	  In view of the fact that $\varepsilon =\t/L_G \ll 1$, and $L_G$ is the smallest geometric scale, the correction $\frac{1}{6} R_{AB}X^AX^B $ is of order $\varepsilon^2$.
 	 
 	  The volumes $V(X)$ and $V_+(X)$  can similarly be expanded around the flat ones in the neighborhood $\mathcal{U}$. Using different explicit geometric setups; in particular different choices for the hypersurface $\Sigma$,  Barton et al suggested the following general expansion for the volumes
 	  
 	  \bea\label{volumec}
 	  V(X_p)&=&\tilde{V}(X_p)\big[ 1+\sum_i \mathcal{G}_i(q)f_i(X_p)+O(\varepsilon^2)\big] \ ,
 	  \nonumber \\
 	   V(X_p)_+&=&\tilde{V}_+(X_p)\big[ 1+\sum_i \mathcal{G}_i(q)f_{+,i}(X_p)+O(\epsilon^2)\big]\ ,
 	  \eea

 	where $\tilde{V}(X_p)$ and $\tilde{V}_+(X_p)$ are the volumes from the all-flat case discussed previously.  $f_i(X_p)$ and $f_{+,i}(X_p)$ are functions of length dimension $L$.
 	
 	Using equations (\ref{metrixexp}) and (\ref{volumec}) the following expansion for $I^{(d)}(q;l_c,\t)$, it is easy  to obtain
 	
 \begin{eqnarray}\label{Iexpan}
 	I^{(d)}(q;l_c,\t)&=&l^{-(d+2)}\int_{0}^{\t} dX^0e^{-\rho_c \tilde{V}(X^0)} \bigg\{\int_{-X^0}^{X^0} dX^1 \tilde{V}_+(X)  \nonumber \\
 			&+&\sum_i\mathcal{G}_i(q)\bigg[ \int_{-X^0}^{X^0} dX^1 \tilde{V}_+(X)f_{+,i}(X)  \nonumber \\ 
 			&-& \rho_c\tilde{V}(X^0)\int_{-X^0}^{X^0} dX^1 \tilde{V}_+(X)f_{i}(X)\bigg] +O(\epsilon^2)\bigg\} \ ,
 \end{eqnarray}
 	 
 	where the fact that the flat cone volume $\tilde{V}$ only depends on $X^0$ was used, and the subscript $p$ from the   coordinates $X^A$ have been removed.
 	
 	 The integral in the first line is just the flat contribution $ I^{(d,flat) (l)}$ given by (\ref{I3}) up to a difference which vanishes exponentially fast in the limit $l_c\rightarrow 0$. 
 	 
 	 By dimensional argument and using Watson's lemma again,  the expression in square bracket of (\ref{Iexpan}) can be shown to evaluate to a term of the form $Cl_c$, for some constant $C$, as $l_c\rightarrow 0$. Similarly, the $O(\varepsilon^2)$ corrections tend to a function of order $O(l^2)$. Therefore the expansion of (\ref{Iexp}) follows.
 	 
 	The constants $a^{(d)}$ are given by their flat values, $ I^{(d,flat)}=a^{(d)} $. The explicit form of the constants $b_i^{(d)}$ can determined once a geometric setup is chosen. For instance Barton et al have explicitly evaluated these constants for two different geometric setups \cite{Barton:2019okw}.

 	 \subsection{The null hypersurface case} 
 	 
 	 The horizon molecules proposal of Barton et al was  specially  devised to work when hypersurfaces of spacelike nature are considered. However, and as we mentioned earlier, there are  good reasons for requiring  any horizon molecule definition to be also valid in the case of null hypersurfaces.   This issue was  not  raised nor discussed  in \cite{Barton:2019okw}, but  a subsequent recent work by Machet and Wang  addressed this question and investigated in detail the extension of   this definition  to encompass  null hypersurfaces intersecting the horizon.  The goal of the following subsection is to give a concise report of the main results and conclusions of  Machet and Wang.
 	 
 	Let us first give a general look at the problem to see how the success of  Barton et al proposal is tied  to the spacelike nature of the hypersurface.  
 
 	To that end consider the all-flat case of figure 9,  a Rindler horizon in Minlowski space, with $\Sigma$ being a straight null plane.
 As can easily be seen the   region $I^-(\Sigma ) \cap I^+(\mathcal{H})\cap I^+(p_-)$ 
 is unbounded with  infinite volume for any randomly selected point $p_-$, and  the expected number of  such horizon molecules is thus directly derived to zero. Therefore, before any sensible calculation of the expected number is started, one has to first bound this domain, at least for the all-flat case. This can be done by either considering a folded null plane instead of the straight one, or by taking a null hypersurface with different shape like a  downward light-cone. These two configurations  were probed   in \cite{Machet:2020uml} to compute the expected number of horizon molecules, although the motivation there for  bounding the region $I^-(\Sigma ) \cap I^+(\mathcal{H})\cap I^+(p_-)$ was to avoid any   potential IR divergence . 
 
 Let us note that, as we discussed in subsection \ref{spacelike}, one can not approach the null case by invoking a continuity argument, like the one given by equation (\ref{limitnullspace}), by  continuously deforming the spacelike result to obtain that of  a null hypersurface. Therefore the issue can only be settled by explicit calculation.

 	\subsubsection{Rindler horizon in Minkowski spacetime}
 	 
 	 Machet and Wang  applied  Barton et al definition  first to   a Rindler horizon in Minckowski space. As we have already mentioned, in the
 	 %horizon in Minckowski space. ِAs we have already mentioned, in the
 	  all-flat case and due to the unboundedness of the region $I^-(\Sigma ) \cap I^+(\mathcal{H})\cap I^+(p_-)$ 
 	 the expected number of Barton et al horizon molecules is identically zero, therefore for any sensible calculation to get started with such definition and geometric setup one has first to introduce some IR regulator to bound the domain $I^-(\Sigma ) \cap I^+(\mathcal{H})\cap I^+(p_-)$. This can be achieved for instance  by taking $\Sigma$ to be a folded null plane or a downward light-cone. 
 	 
 	 \begin{figure}
 	 	\begin{center}
 	 		\includegraphics[height=2.5in,width=4in,angle=0]{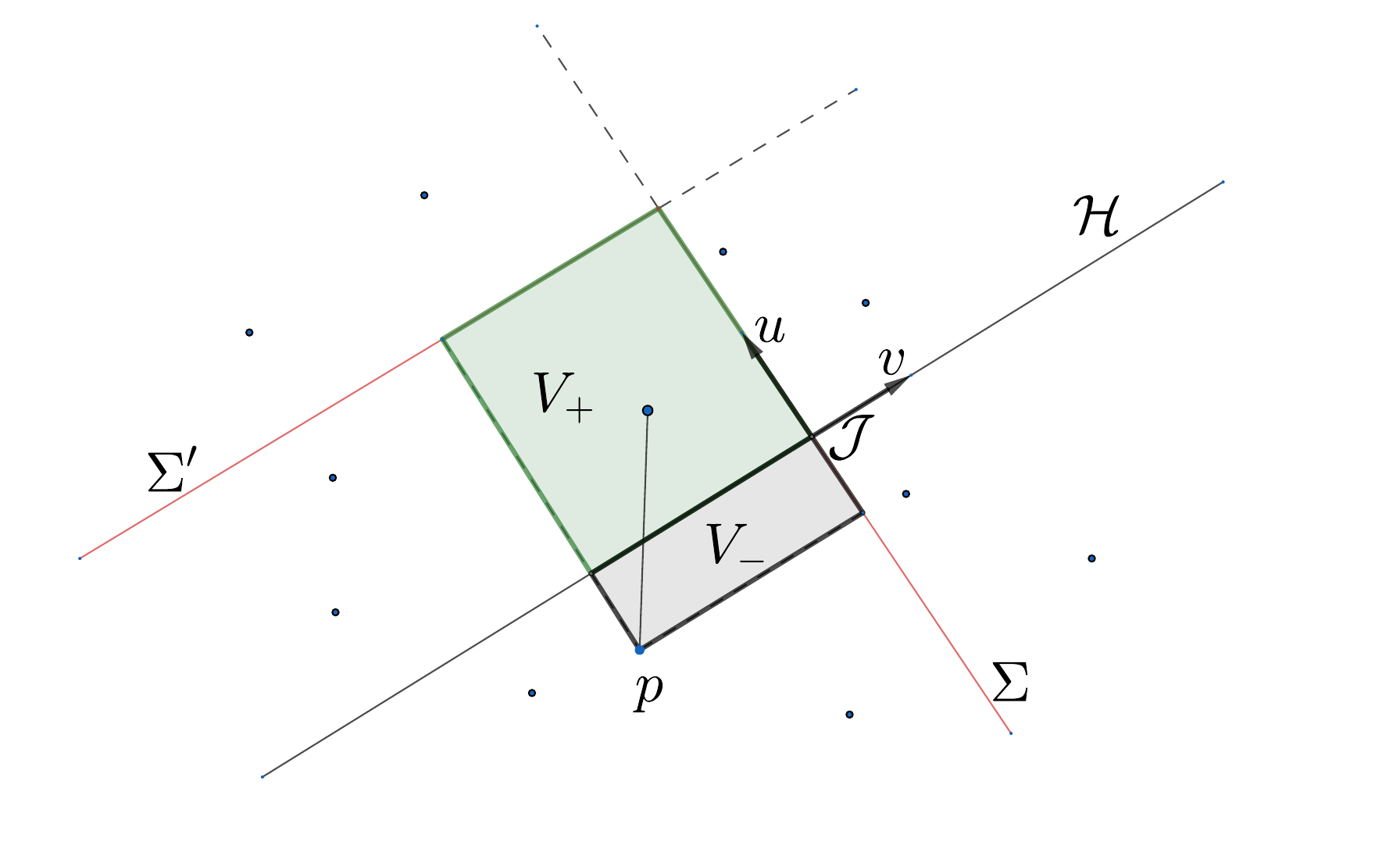}
 	 		\caption{An all-flat setting: a folded null plane  crossing in a  Rindler horizon .}
 	 	\end{center}
 	 \end{figure}

 	 The setup of a folded null plane is depicted in Figure 9.  A global coordinates $(v,u,y^\a)$ is set up, the null hypersurface $\Sigma$ is the $u$-axis, $v=0$ and the horizon $\mathcal{H}$ is the $v$-axis, $u=0$.  Another null hpersurface $\Sigma '$ given by $u=\lambda$ , with $\lambda>0$.  The union $ \Sigma '\cup \Sigma$ is the folded null plane with respect to which the expected number of horzion molecules is to be counted.
 	 
 	 In this setup the volumes $V_+(p)$ and $V(p) $ can explicitly be computed for arbitrary dimension, and  are given by
 	 
 	 $$
 	 V_f(u,v,\lambda)= \a_d (2v(u+\l))^{d/2}, $$
 	 $$
 	  V_{+,f}(u,v,\lambda)=\a_d (2v)^{d/2}\big( (u+\l)^{d/2}-u^{d/2}\big) \ .
 	 $$
 	 
 where $\a_d$ is a constant dependent on the dimension.
 
It follows again that the expected number of horizon molecules can be written as

\be\label{expnullf1}
	  \varrho_c^{\frac{2-d}{d}}<\mathbf{H}_1>= \int_{\mathcal{J}} d^{d-2}y  I^{(d,flat)}_{null}(l_c,\l) \ ,
 	  \ee
 	  
 	 where $I^{(d,flat)}_{null}(l_c,\l)$  is given by
 	 \be
 	 I^{(d,flat)}_{null}(l_c,\l)= l_c^{-(d+2)} \int_{0}^{\infty} dv\int_{0}^{\infty} du V_{+,f}(u,v,\lambda) e^{-\rho_c V_f}  \ .
 	  \ee

 	  Using the explicit formulas for $V_{+,f}$ and $V_f$ one gets
 	  \be
 	  I^{(d,flat)}_{null}(l_c,\l)=\frac{\a_d^{-2/d}}{d} \Gamma [2/d+1] \int_{0}^{\infty} du\big( (u+\l)^{d/2}-u^{d/2}\big) (u+\l)^{-1-d/2} \ .
 	   \ee
 	  It is noticeable that any dependence on the discreteness scale has disappeared and therefore, by dimensional analysis, $I^{(d,flat)}_{null}(l_c,\l)$  should be independent of $\l$. The above integral can be evaluated and one obtains
 	  \be
 	  I^{(d,flat)}_{null}(l_c,\l):= I^{(d,flat)}_{null}=a^{(d,flat)}_{null}= \frac{\a_d^{-2/d}}{d} \Gamma [2/d+1]H_{d/2} \ ,
 	   \ee
 	   
 	   where $H_k$ is the $k^{th}$ harmonic number.
 	  
 	  Actually, Machet and Wang also carried out the calculation for $n$-molecule and obtained a formula which can be exactly evaluated for each $n$.
 	  
 	  It follows then
 	  \be\label{arealawnull}
 	  \varrho_c^{\frac{2-d}{d}}<\mathbf{H}_1>=a^{(d,flat)}_{null}\int_{\mathcal{J}_t} dV_{\mathcal{J}} \ .
 	  \ee
 	  
 	  Some comments on this result are in order.
 	  
 	  It is first interesting to note that the resulting constants $a^{(d,flat)}_{null}$ are different from the constants $a^{(d,flat)}$ obtained in the spacelike case, as can be checked by substituting for particular values of $d$.

 	  Moreover, the final result is independent of the position of the $\Sigma '$, the parameter $\l$ . Therefore one may take the IR regulator to infinity without changing the result, hence going back to null plane case, and this goes in contradiction with the fact that if one started with a null plane the expected number of horizon molecules would be identically zero. Again, we see that this horizon molecules counting is sensitive to how some limits are taken.
 	  
 	 The independence of $a^{(d,flat)} $ from $l_c$ is actually related to its independence of $\l$ and flat setup used to do the counting. As it was pointed out in  \cite{Machet:2020uml}, because of Lorentz invariance of the counting and the fact that one can always boost the system in the $u$ direction to pull the surfrace $\Sigma '$ arbitrarily close to  $\mathcal{H}$ the result should not depend on $\l$. If $\mathcal{J}$ has no geometrical quantity associated to it, e.g intrinsic curvature, then $l_c$ has no length  scale to couple with and therefore $a^{(d,flat)} $ can only be a pure number. Similarly to the spacelike case, to establish the area law in the all-flat setup using this counting the continuum limit plays no role, equation (\ref{arealawnull}) is valid for  any finite $\varrho_c$.

 	  Another setup probed in \cite{Machet:2020uml} was again a Rindler horizon in Minkowski space but with a null hypersurface $\Sigma$ having a different shape, namely a downward light-cone. 
 	 
 	 \begin{figure}
 	 \begin{center}
 	 	\includegraphics[height=2.5in,width=4in,angle=0]{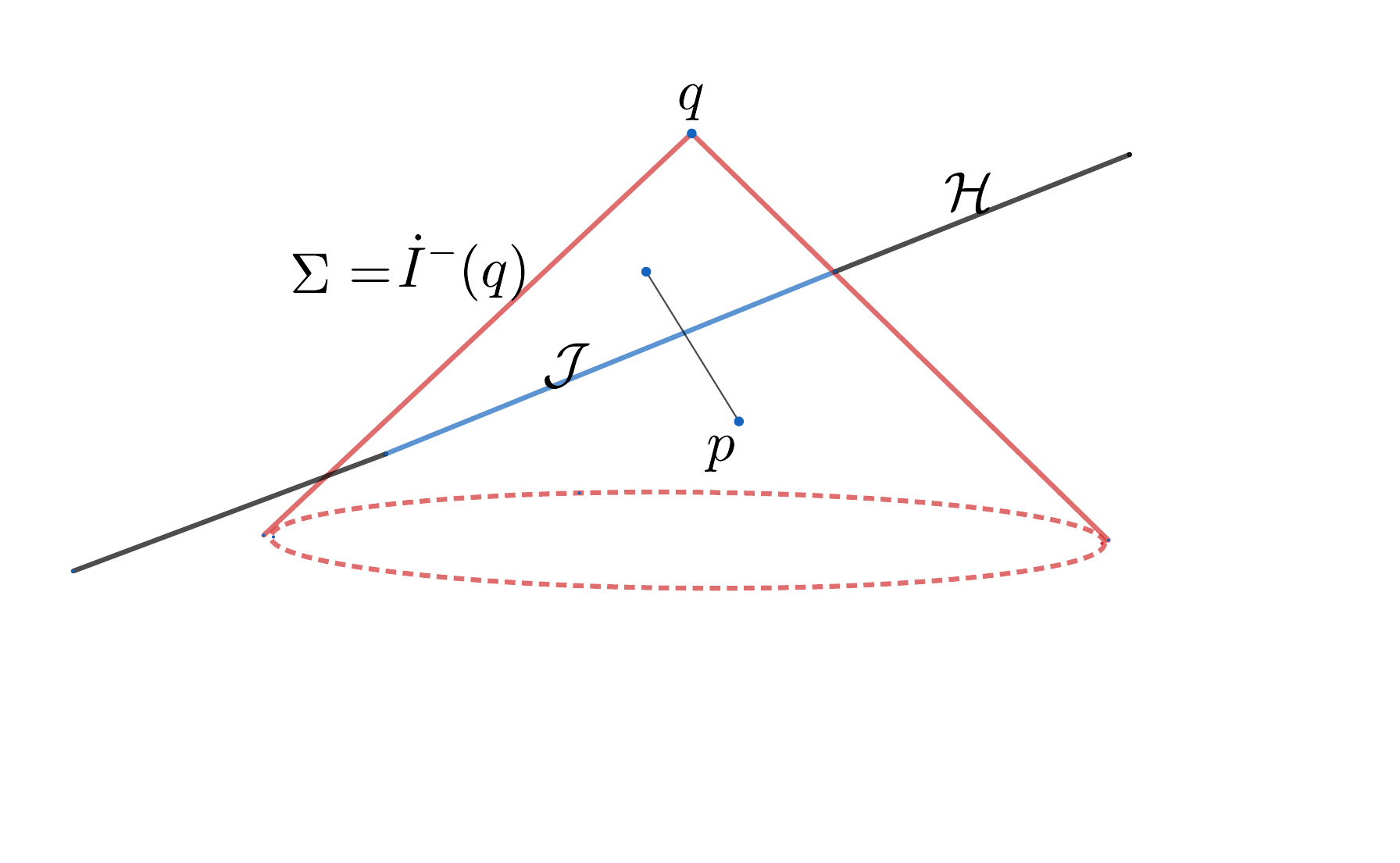}
 	 	\caption{A downward light-cone intersecting a Rindler  horizon. }
 	 \end{center}
  \end{figure}

 	  The downward light-cone $\Sigma$  is defined to be the boundary of the causal past of a point $q \in \mathcal{H} $ , i.e $\dot{I}^-(q)$, Figure 10.
 	  
 	  The volumes $V_+$ and $V(p)$ are now given by
 	  
 	  $$
 	  V_{+,q}+(p):= \text{vol}(I(p,q)\cap I^+(\mathcal{H})), ~~~~ V_q(p):= \text{vol}(I(p,q)) \ .
 	  	$$
 	  	
 	  	Machet and Wang could compute these volumes explicitly  in $d=4$, in terms of the null coordinates of $p$ and the affine distance between the horizon and the point $q$, and  obtained a general  formula for $I^{(4,flat)}_{cone}$ which reduces for $n=1$ to 
 	  	
 	  	\be
 	  	I^{(4,flat)}_{cone} \frac{1}{12}\sqrt{\frac{3}{2}}\bigg(\frac{27}{4}-{}_2F_1(1,-1,-2;\frac{3}{2})  \bigg)= a^{(4)}  \ .
 	  	\ee
 	 
 	  We notice that $I^{(4,flat)}_{cone}$  turns out to be again independent of the discreteness scale and of the other length scale provided by the affine distance between the horizon and the point $q$. The explanation of this independence  is similar to the folded null plane.
 	  
 	  It is noteworthy that here too the constant $a^{(4)} $ is different from its counterpart in the spacelike hypersurface.
 	  
 	  If one insists on the necessity that the null and spacelike hypersurface must give the same proportionality constant to the horzion area, and take it as a sanity condition of any horizon molecules proposal, then we see that  the horizon molecules definition introduced by Barton et al does not meet this requirement.
 
A final remark about the flat counting with a null hypersurface  is that it is free from any IR divergences. This IR divergences could have arisen from points $p_-$ arbitrarily close to the $\Sigma$ and to the far past of $\mathcal{J}$, e.g with $v\sim 0$ , $u\rightarrow -\infty$ and   volume $V_f$ close to zero in the folded plane case, hence not exponentially  suppressed. It is not actually difficult to see how this IR divergence is cured within this horizon molecule counting. The requirement that $p_-$ is max-but-$n$ (at least $n=1$) bounds it away from  $\Sigma$, this is realized in the general integral formula of $<\mathbf{H}_1>$  by multiplying the exponential term, $e^{-\varrho_c V}$, by the extra volume term $V_{+,f}$, or $V_+^n$ for $n$-molecule, which  has first to vanish before  $V_f$ approaches zero. Therefore, the term $V_{+,f}$ accompanying the exponential   kills off this IR.

 	  \subsubsection{Curved case}  
 	  To investigate the curved case with a null hypersurface the author in \cite{Machet:2020uml} took a  path similar in spirit to that taken by Barton et al, but now by setting up a local  Guassian Null Corrdinates (GNC) adapted  to the study of the null hypersurface case. 
 	  
 	 For  folded null planes, a local coordinates $(v, u, y^\a) $ is constructed in a tubular neighborhood $ \mathcal{N} \supset \mathcal{J}$. A region $\mathcal{R}_\Lambda$ analogue to $\mathcal{R}_t$ in spacelike hypersurface case is defined as follows (in  time-reversed  corrdinates )

 	 \be
 	 \mathcal{R}_\Lambda:= \{ p\in I^+(\mathcal{J})\cap \mathcal{N} | 0<v(p)<\Lambda , 0<u(p)<\Lambda \} \ ,
 	  \ee 
 	 where $\Lambda$ is an intermediate scale between the discreteness and the geometric length of the setting, i.e $l_c\ll \Lambda \ll L_G$. For the argument used in the spacelike hypersurface case to carry over to the null setting one has to show that the rescaled expected number of horizon molecules can be reduced to a local integral on  $\mathcal{R}_\Lambda $
 	 
 	 \be\label{nulltruncated}
 	  \varrho_c^{\frac{2-d}{d}}<\mathbf{H}>=\varrho_c^{\frac{2-d}{d}+2} \int_{\mathcal{R}_\Lambda}   V_+(p,\l) e^{-\varrho_c V(p,\l)} dV_p +\cdots  \ ,
 	  \ee
	  where $``....."$ refer to terms decaying exponentially fast as we go to the limit $l_c \rightarrow 0$, and $\l$ is the parameter of the folding hypersurface $\Sigma '$.

	Based on the discussion of the flat setting, it is not difficult to argue that the above local  expansion can not in general be true.  This can be seen by first noticing that  although the region ( in future-past swapped setting) $v(p)> \Lambda$ poses no problem for all values of $u(p)$, as $\varrho_c V \gg 1$, so that all contributions from this region will be exponentially suppressed in the continuum limit. However, when $u(p) >\Lambda $ this argument fails,  contributions coming from   points $p$  close to $u$-axis, with $v(p)\sim 0$,  are not exponentially suppressed, as the volume $V(p)$ is now close to zero for values of $u$ arbitrarily far in the future (or the past in the original setting ) of $\mathcal{J}$. Thus, contributions coming from far away along the past light-cone of the intersection hypersurface can not be neglected. It follows then that the above local integral can not in general count for the dominant contributions to the expected number of horizon molecules in the continuum limit.  Therefore Machet and Wang concluded that this failure is a first indication that the proposal of Barton et al to count horizon molecules with a null hypersurface is a flawed way to define entropy on causal set setting.
	
	 Machet and Wang further argued that a truncated (by hand) local integral in the form (\ref{nulltruncated}), in which contributions from the far past of the inetsection are excluded, yields a small $l_c$ expansion of the following form for $ I^{(d)}(y;\l_c, \l, \Lambda)$ 
	
	\be
	I^{(d)}(y;l_c, \l, \Lambda)=a^{(d)}+\sum_i  c_i^{(d)}\mathcal{F}_i(y,\l,\Lambda)+ l_c \sum_i b_i^{(d)} \mathcal{G}_i(y,\l,\Lambda) + O(l_c^2) \ ,
	\ee
	  where $a^{(d)}$, $b_i^{(d)}$ and  $c_i^{(d)}$ are dimensionless constants dependent on $d$. The set ${\mathcal{G}_i(y,\l,\Lambda)}$ is a set of mutually independent geometrical scalars of length $L^{-1}$  evaluated along a geodesic segment. The extra set ${\mathcal{F}_i(y,\l,\Lambda)}$  is set of independent geometrical  dimensionless scalar evaluated along a geodesic segment.

	 The presence of extra terms ${\mathcal{F}_i(y,\l,\Lambda)}$ carrying geometrical information evaluated along   the geodesic segment which, in contrast to the spacelike hypersurface case, survive in the limit $l_c \rightarrow 0$ can be given the  heuristic explanation based on the above discussion and dimensional analysis.
	
	To further support their claim Machet and Wang worked out an explicit geometrical setting in which $\Sigma$ is a downward light-cone, the past-pointing light-cone of a point $q$ which is of affine distance $\l$ away from $\mathcal{J}$. Within the relevant region, in which the $V$ and $V_+$ tend to a skinny causal interval or diamond, a Null Fermi Normal coordinates  system $(b,u,y^\a)$  is set up. Figure 11  gives a sketch of the coordinates system. 
	
	\begin{figure}
	\begin{center}
		\includegraphics[height=2.5in,width=4in,angle=0]{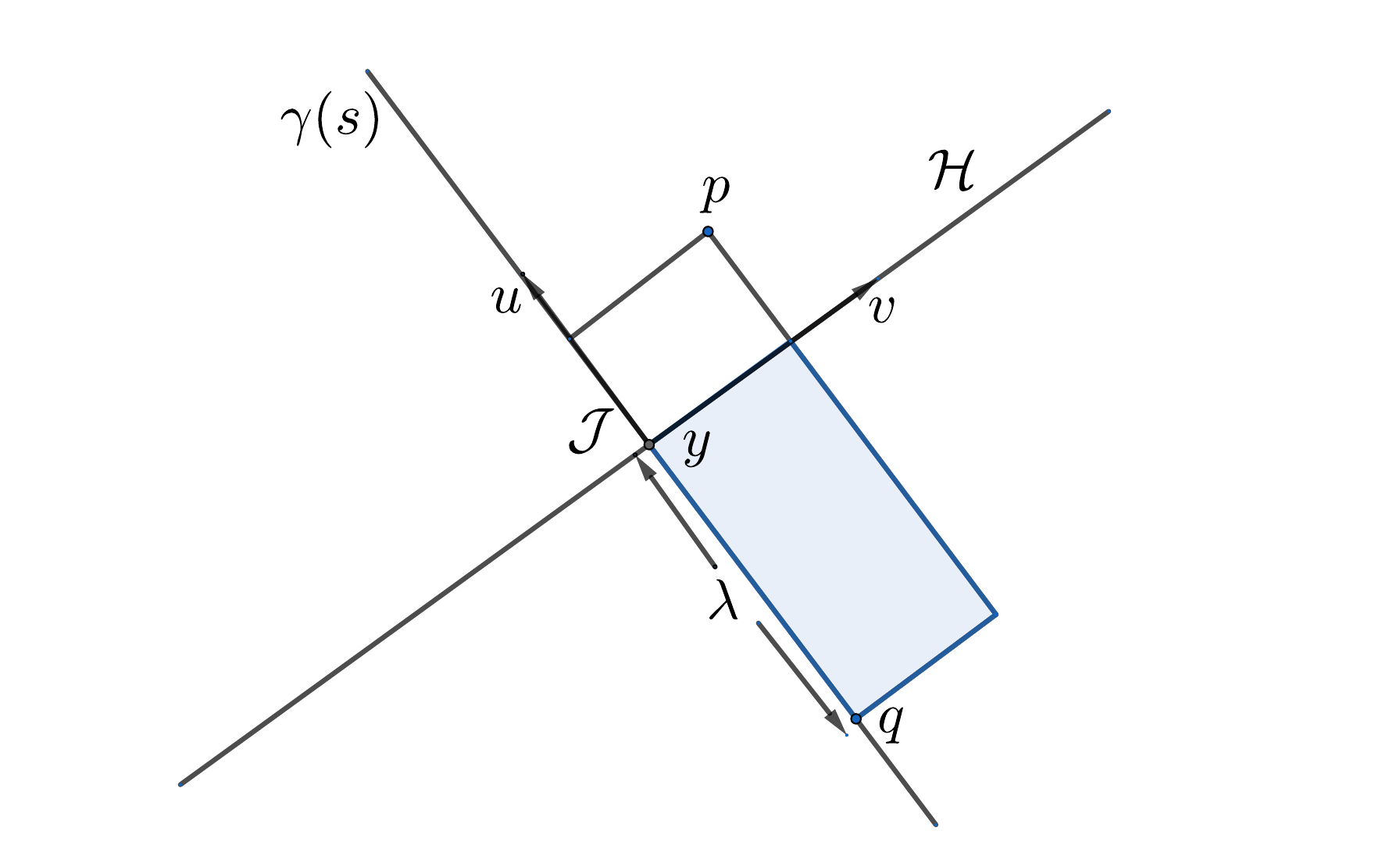}
		\caption{A sketch of the coordinates system in the relevant skinny diamond setup. $V_+$ is shaded.}
	\end{center}
\end{figure}

	Both $V$ and $V_+$ can be expanded around the flat ones and  admit the following expansions in $d=4$,
	\be
	V^{(4)}(u,v,y;\l)=	V_f^{(4)}(u,v,y;\l)\bigg( 1+ F(\l, u)+\mathcal{O} ((u+\l)^3,v^3)\bigg) \ ,
	\ee
	
	\be
	V^{(4)}(u,v,y;\l)=	V_f^{(4)}(u,v,y;\l)\bigg( 1+ \tilde{F}(\l, u)+ \mathcal{O}((u+\l)^3,v^3)\bigg) \ ,
	\ee
	where $F(\l, u)$ and $\tilde{F}(\l, u))$  are two dimensionless function involving the integration over Ricci tensor along the $u$ direction.  The assumption here is of course that $\l$ is small relative to the local curvature scales,  $u$ is to be cutoff at distance $\L$  small compared to the local curvature scales.
	
	Under the above assumptions along with an extra assumption about the Ricci tensor (an assumption generic enough to support the claim) it was possible to show that $ I^{(4)}(y;\l_c, \l, \Lambda)$  admits the following continuum limit
	
	\be
\lim _{l_c \rightarrow 0}I^{(4)}(y;l_c, \l, \Lambda)=a^{(4)}+\mathcal{R}(y)c^{(4)}(\l, \L)+\cdots \ .
\ee
 One can see that the limit is not local to the intersection $\mathcal{J}$, and the area law is distorted. Machet and Wang then concluded that the horizon molecules proposal of Barton et al does not yield a well behaved area law for when evaluated on a null hypersurface intersecting  the horizon.
 
 Now, whether the above argument is conclusive or just an artifact  of the  limitation of the expansion adopted by Machet and Wang, which  relies on an ad hoc truncation by hand of the integral $I^{(4)}$ and  certainly neglecting relevant contributions, remains in our view an unsettled issue. One, for instance, cannot exclude the possibility that in a realistic BH model the area law  might be restored. A hint for this possibility is offered by the links counting in 2-dimensional reduced Schwarzschild BH discussed in section \ref{link}, where the area law is established in the null surface case in a quite subtle way, and at  the end the dominant contribution  turns out to plainly comes from the near horizon links.
 
	\subsection{Discussion and outlook}
	
	In this   survey, we have tried to take the reader  through the different attempts to identify the right horizon molecules that would give a good kinematical account for  BH entropy  within the causal set approach . Despite  of the fact that there have only been  few scattered efforts and 
	 practitioners who have devoted their time to this issue, it is undeniable that some progress has been made along different directions. The  simplicity, the success in 2-dimensions and the failure in higher dimensions of the causal links proposal stimulated further investigations and proposals based on triplets, and recently has sparked  interest in the subject.
	 
	 The  early proposal based on causal links gave some  promising results in two dimensions, some of which may seem surprising . Prominent among them is the fact that one finds a universal answer  which took the same value in two quite different black hole backgrounds, that of  equilibrium and non-equilibrium cases, and that the bulk of the links always  reside in the close proximity to the horizon, meaning that the result is controlled by the near horizon geometry . However, a seemingly surprising result is the that this value remains finite even in the continuum limit where the fundamental length $l_c$ is sent to zero. In this sense, the replacement of continuous spacetime by a causal set may appear in two dimensions as more of a regularization device then something fundamental. Whether this has any deeper meaning, or whether it might be related to some of the other properties that both quantum field theory and quantum gravity possess in two  dimensions \cite{Fiola:1994ir}, that remains an open question. Of note is also the fact that the causal set approach to quantum gravity has been unique in attempting to account for the statistical mechanics of the non-equilibrium horizon.
	 
	 It must be added that the above features of the causal links counting are  shared by   the triplet proposals, i.e. universality of the result and the finiteness in the continuum limit in two dimensions.  
	 
	 Two  questions  about the links and triplets proposals have so far remained open. The first is to find a way to decide  whether the spacelike hypersurface gives the same result as the null one in two dimensions, regardless of the validity of these proposals beyond two dimension. But as we early mentioned this a mere mathematical curiosity with  little, if any, physical relevance.  The second is of importance and concerns the triplet proposal. Although Marr \cite{Sarah} argued that the $\Lambda$-triplet would suffer from IR divergences  in higher dimensions, the issue is unsettled for $l$ and $z$-triplets. Therefore, it would be  interesting exercise  to investigate this triplets counting at least in three  dimensional flat setting. However, as we already mentioned, going beyond two dimensions would make the calculation cumbersome, but if enough time and effort is devoted to this problem, some   approximation  methods could possibly  be devised to extract the leading contributions or settle   the divergence issue. Of course, one could also use numerical methods to approach this counting.

	    Unlike the links and the triplets attempts, the Barton et al proposal has succeeded in giving  an expected number of horizon molecules proportional to the area of the horizon intersecting a spacelike hypersurface, for almost all reasonable geometrical settings. However, this proposal has some drawbacks. Firstly, it seems to be inherently discontinuous as one moves from the spacelike hypersurfaces to null ones, giving two different values, i.e different proportionality constants.  Moreover, if one accepts the  conclusion of Machet and Wang \cite{Machet:2020uml}, in  a curved geometrical background and for a null hypersurface, the continuum limit of the expected number of  Barton et al horizon molecules is not local to the intersection of the horizon and the hypersurface, yielding to an ill-behaved area law. Nonetheless, if one sticks with spacelike hypersurfaces, the Barton et al counting provides a good measure for the area of the intersection of the horizon (a null surface) and spacelike hypersurface, which is a promising aspect of such horizon molecules proposal.

	    Another weakness of Barton el al definition,  in our view, is that it lacks the physical intuitive and heuristic picture  shared by the links and the triplet (and the diamond) proposals. The elements that underpinned the Barton et al horizon molecules are points exterior to the black hole and to the past of the hypersurface, with no reference to the future of the hypersurface,  hence it would be hard to view such molecules as heuristically producing entanglement during the course of the causal set growth (or time  development).

	    Finally, the common weakness of all proposals discussed in this review, of course,  is that they remain at purely kinematical level, and even if a fully successful  kinematical identification of the horizon molecules is achieved,  no successful proposal can be substantiated or refuted before we possess a fully quantum dynamics of causets. And despite a very  important step  made by Rideout and Sorkin in developing classical stochastic dynamics for causets, \emph{classical sequential growth } models  \cite{Rideout:1999ub},  building  a viable quantum sequential
	    growth  dynamics has so remained challenging \cite{Sorkin:2011sp, Martin:2004xi, Surya:2019ndm}.        
	 %As seen in the case  of space like case, in the continuum limit the integrand of $I^{(d)}$ is non-negligible only in the neighborhood of $y$ , where the volume $V$ shrinks to zero. Therefore the small $l_c$ expansion is dependent only on the geometric scalars evaluated at $y$. There are no other scales than $l_c$ which can pair with the curvature tensors to form dimen
	 \hfill \break
	 
\textbf{Acknowledgments}

While  writing this review I have benefited from several discussions with Fay Dowker and Ludovico Machet, I am grateful to them for answering my questions and discussing several issues regarding their works.  

\bibliographystyle{JHEP}
\bibliography{document.bib}
\end{document}